\documentclass[prd,twocolumn,superscriptaddress]{revtex4-1}
\usepackage{amsmath}
\usepackage{amssymb}
\usepackage{graphicx}
\usepackage[english]{babel}
\usepackage{blindtext}
\usepackage{color}
\usepackage[dvipsnames]{xcolor}
\usepackage{graphicx}
\usepackage{dcolumn}
\usepackage{bm}
\usepackage{hyperref}
\usepackage{mathtools}
\usepackage[normalem]{ulem}

\usepackage{verbatim}
\usepackage{mathrsfs}
\usepackage{amsmath}

\newcommand{\ket}[1]{\left| {#1} \right\rangle}
\newcommand{\bra}[1]{\left\langle {#1} \right|}

\newcommand{\proj}[2]{\left| {#1} \right\rangle\!\left\langle {#2} \right|}
\newcommand{\ii}{\mathrm{i}}

\renewcommand{\d}{ \mathrm{d}}
\newcommand{\tr}{\operatorname{Tr}}

\newcommand{\op}{\hat}
\newcommand{\fhi}{\op{\phi}}

\newcommand{\be}[1]{\begin{equation}\label{#1}}
\newcommand{\ee}{\end{equation}}
\definecolor{pink}{rgb}{0.8, 0.1, 0.5}
\renewcommand{\aa}{{\textsc{a}}}
\newcommand{\bb}{{\textsc{b}}}
\renewcommand{\Im}{\textrm{Im}}

\begin{document}

\title{Entanglement harvesting and divergences in quadratic Unruh-DeWitt detectors pairs}

\author{Allison Sachs}
\email{asachs@uwaterloo.ca}
\affiliation{Institute for Quantum Computing, University of Waterloo, Waterloo, ON, N2L 3G1, Canada}
\affiliation{Dept. Physics and Astronomy, University of Waterloo, Waterloo, ON, N2L 3G1, Canada}

\author{Robert B. Mann}
\email{rbmann@uwaterloo.ca}
\affiliation{Institute for Quantum Computing, University of Waterloo, Waterloo, ON, N2L 3G1, Canada}
\affiliation{Dept. Physics and Astronomy, University of Waterloo, Waterloo, ON, N2L 3G1, Canada}
\affiliation{Perimeter Institute for Theoretical Physics, Waterloo, ON, N2L 2Y5, Canada}

\author{Eduardo Mart\'in-Mart\'inez}
\email{emartinmartinez@uwaterloo.ca}
\affiliation{Institute for Quantum Computing, University of Waterloo, Waterloo, ON, N2L 3G1, Canada}
\affiliation{Dept. Applied Math., University of Waterloo, Waterloo, ON, N2L 3G1, Canada}
\affiliation{Perimeter Institute for Theoretical Physics, Waterloo, ON, N2L 2Y5, Canada}

\date{\today}

\begin{abstract}
We analyze correlations between pairs of particle detectors quadratically coupled to a real scalar field.  We find that, while a single quadratically coupled detector presents no divergences, when one considers pairs of detectors there emerge unanticipated persistent divergences (not regularizable via smooth switching or smearing) in the entanglement they acquire from the field.  We have characterized such divergences, discussed whether a suitable regularization can allow for fair comparison of the entanglement harvesting ability of the quadratic and the linear couplings, and finally we have found a UV-safe quantifier of harvested correlations. Our results are relevant to  future studies of the entanglement structure of the fermionic vacuum.

\end{abstract}

\pacs{Valid PACS appear here}
\maketitle

\section{Introduction}

The vacuum state of a quantum field displays quantum correlations between observables defined in spacelike separated regions \cite{summers:1985aa,summers:1987aa}. This vacuum entanglement has been studied in quantum foundations, and has found a variety of applications such as quantum energy teleportation \cite{Hotta:2009aa,Hotta:2010aa}, the black hole information loss problem \cite{Susskind:1993aa} and firewalls, along with black hole complementarity \cite{Almheiri:2013aa, Braunstein:2013aa}.

In a phenomenon called \textit{entanglement harvesting} \cite{Salton:2015aa}, correlations in a quantum field (such as the electromagnetic field) can be swapped to particle detectors (such as atoms or qubits). This is possible even when the particle detectors remain spacelike separated throughout the duration of their interaction with the field. This was first shown by Valentini \cite{Valentini:1991aa} and later by Reznik et al. \cite{Reznik:2003aa,Reznik:2005aa}. 

Since then, entanglement harvesting has been  shown to be sensitive to the background geometry of spacetime \cite{Steeg:2009aa,Edu:2012aa,Edu:2014aa}, as well as the topology \cite{Edu:2016aa}. Additionally, it has been shown that entanglement harvesting can be done sustainably and distilled into Bell pairs in a process called \textit{entanglement farming} \cite{Edu:2013aa}, a protocol that can be adapted to create a quantum seismometer \cite{EricBrown:2014aa}. Entanglement harvesting has also been studied in detail in timelike separation contexts \cite{Ralph:2012aa,Ralph:2011aa} with implementation proposals in different testbeds from quantum key distribution based on homodyne detection \cite{Ralph:2015aa} to strongly coupled superconducting qubits \cite{Sabin:2012aa}. 

To model the entanglement-swapping interaction between the detectors and field, the Unruh-DeWitt detector model is used. This  model utilizes a first-quantized system (called a detector) linearly coupled to a scalar bosonic field. While most of our knowledge of entanglement harvesting has been gleaned from this setup \cite{Steeg:2009aa,Edu:2012aa,Edu:2013aa,Edu:2014aa,EricBrown:2014aa,Edu:2016aa,Pozas:2017aa}, there has been some exploration of more realistic models such as electromagnetic coupling of atoms \cite{Pozas:2016aa}. All these studies, however, analyzed entanglement harvesting form bosonic fields.

It is known from fundamental studies that there are differences between the entanglement structure of the vacuum of fermionic and bosonic fields \cite{PhysRevA.74.032326,PhysRevA.80.042318,PhysRevA.81.032320,PhysRevA.81.052305,PhysRevD.82.045030,PhysRevA.82.042332,PhysRevD.82.064006,PhysRevA.83.052306,Montero:2011aa,PhysRevA.84.012337,Montero:2011ab,PhysRevA.84.042320,PhysRevA.84.062111,PhysRevA.85.016301,PhysRevA.85.016302,PhysRevD.85.025012,PhysRevA.85.024301,PhysRevA.87.022338}. However, a study of entanglement harvesting in fermionic setups has never been performed. A study of detector-based entanglement harvesting from a fermionic vacuum  could resolve ambiguities in defining entanglement measures between disjoint regions of a fermionic field \cite{Montero:2011aa,Montero:2011aa,PhysRevA.84.042320,PhysRevA.84.062111,PhysRevA.85.016301,PhysRevA.85.016302,PhysRevD.85.025012,PhysRevA.85.024301,PhysRevA.87.022338}. The reasons why this has not been done can be traced back to fundamental difficulties associated with particle detector models for fermionic fields.

To analyze fermionic fields form the perspective of localized particle detectors  a detector model was introduced
that consisted of a cavity coupled to a fermionic field \cite{Iyer1980}, much like Unruh's original detector was a cavity coupled to a bosonic field \cite{unruh:1976aa}. Later, Takagi introduced an UDW-like model for fermionic fields \cite{Takagi:1985aa,Takagi:1986aa}, wherein a two level system coupled quadratically to a fermionic field,
\begin{align} 
H_\textsc{f} \propto \hat\mu\overline\Psi\Psi.\label{quadferm1}
\end{align}
However, this model contained persistent divergences that could not be regularized by an appropriate choice of switching and smearing functions, as it was found in \cite{hummer}. Thus, these investigations restricted themselves to studying transition rates instead of transition probabilities. To track down the origin of these divergences in Takagi's fermionic detector, H\"umer et al. studied three types of quadratically coupled UDW-like detectors  \cite{hummer}. They concluded that these divergences are mainly due to the detector coupling quadratically to the field instead of resulting from the analytic structure (spinor vs. scalar) or statistics (fermionic vs. bosonic) of the fields involved. Persistent divergences in the single-detector vacuum excitation probabilities (VEP) of quadratic UDW detectors were found to be renormalizable by the same techniques used in QED \cite{hummer}, i.e. normal ordering the interaction Hamiltonian.

The analysis in \cite{hummer}, however, was limited to a single detector coupled to the field. It is therefore natural to  extend the studies of the quadratic coupling to fermions to settings of many detectors to explore, for example, entanglement harvesting from a fermionic vacuum.  Nevertheless, before moving to the fermionic coupling, it is important to understand how entanglement harvesting works for detectors quadratically coupled to a bosonic field, so as to determine how much of any new phenomenology would be due to the fermionic nature of the field and how much of it is due to the quadratic nature of the coupling.

In this paper we extend the notion of entanglement harvesting to a detector coupled \textit{quadratically} to a bosonic field.  While this does not answer open questions about entanglement ambiguities in fermionic fields as posed in \cite{Montero:2011aa}, it does shed light on differences between linearly and quadratically coupled detectors, which is one prominent difference between the bosonic and fermionic UDW models.  A fermionic UDW model could resolve ambiguities in defining entanglement measures between disjoint regions of a fermionic field \cite{Montero:2011aa} and to this end it is important to understand how it is distinguished from its bosonic counterpart.

We find that, despite the finite renormalized single detector vacuum excitation probability, the two-detector density matrix  remarkably contains persistent divergences at leading order in perturbation theory. These divergences cannot be regularized by means of a smooth switching function or spatial profile, nor are they renormalized by the techniques used in \cite{hummer}. Instead, one must introduce additional means of regularization (e.g., a soft UV cutoff). It is interesting to note that these divergences appear only in the non-local contributions to the density matrix. As a result the entanglement harvested by detectors quadratically coupled to bosonic fields  is sensitive to the choice of UV cutoff and may require further regularization.  

To tackle the problem of entanglement harvesting with a pair of quadratically coupled detectors, we will follow two different avenues: a) We will analyze the nature and strength of the divergences in 3+1D flat spacetime, analyzing possible physically motivated regularization scales in entanglement harvesting and b) we will propose a measure of correlations between the detectors that are divergence free and use it to further our knowledge of the differences between the use of linear and quadratic couplings of particle detectors to study the entanglement structure of quantum fields.

This paper is organized as follows. In Section \ref{UDWdetector} we introduce the linear and quadratic UDW detector models and examine in detail their time evolution. In Section \ref{singledetector} we provide an overview of the single UDW detector, both quadratically and linearly coupled to a scalar bosonic field. Section \ref{twodetector} analyzes the two-detector entanglement-harvesting set up; we show in detail how persistent divergences emerge in the non-local terms of the quadratically coupled two-detector system. In section \ref{harvesting} we compare the entanglement harvesting capabilities for the linear and quadratic models, first looking at entanglement harvesting under suitable UV-regularization and then studying a divergence free quantifier of harvested correlations from the field: the mutual information. We present our conclusions in section \ref{ap:convolution}.

\section{Time evolution of linear and quadratic detector models 
\label{UDWdetector}}

Let us introduce the two different detector models that we will analyze and compare in this paper. First, let us consider the well-known UDW detector model. This model was first introduced as an operational way to study the particle content of a bosonic quantum field \cite{unruh:1976aa,dewitt:1979aa}. It consists of a two-level quantum emitter (detector) coupled linearly to a scalar quantum field along its worldline.

For a single inertial detector (labelled A) in flat spacetime, the UDW interaction Hamiltonian in the interaction picture is given by
\begin{align} 
    \op H_{\op \phi} (t) = \lambda_{\aa} \chi_{\aa}(t-t_{\aa})  \op \mu_{\aa}(t) \! \int  \d^n \bm{x}  F_\aa (\bm{x}-\bm{x}_{\aa})  \op{\phi}(\bm{x},t).
\label{hamlin}
\end{align}
Here, the monopole moment $\op \mu_\aa(t)=\op \sigma^+_\aa e^{\ii \Omega_\aa t}+\op \sigma^-_\aa e^{-\ii \Omega_\aa t}$ represents the two-level internal degree of freedom of the detector, which couples linearly to a real massless scalar field $\fhi(\bm{x},t)$. $0\le\chi_\aa(t)\le 1$ is the switching function that controls the time-dependence of the coupling of  strength $\lambda_A$.   The spatial profile $F(\bm{x})$ carries information about the shape and size of the detector. The case of the point-like detector, commonly considered in the literature, is a particular case of \eqref{hamlin} where the smearing function is a delta distribution, $F_\aa(\bm{x})=\delta(\bm{x})$.

Modifications of this model where the detector is coupled quadratically to the field \cite{Hinton}  allow one to compare on equal footing the response of bosonic and fermionic detectors (the latter necessarily being quadratic \cite{Takagi:1986aa}). These models have been recently analyzed in detail in  \cite{hummer}.  The interaction Hamiltonian for a quadratically coupled UDW detector is given by 
\begin{align} 
    \op{H}_{\op{\phi}^2} (t) =  \lambda_{\aa} \chi_{\aa}(t-t_{\aa})  \op{\mu}_{\aa}(t) \! \int  \d^n \bm{x}\,  F_\aa(\bm{x}-\bm{x}_{\aa})  :\!\op{\phi}^2(\bm{x},t)\!:,\label{hamquad}
\end{align}
where $\phi^2(\bm{x},t)$ has been normal-ordered as prescribed by the analysis in \cite{hummer}.

It is convenient at this point to define two different types of UV divergences that particle detector models may present. A \textit{regularizable} divergence is one that can be removed by use of a smooth switching function and/or a smooth spatial profile (see, e.g., \cite{Jorma:2008aa,Jorma:2006aa,satz:2007aa}). A \textit{persistent} divergence is one that remains even with smooth switching and smearing functions (such as the divergences renormalized in \cite{hummer}).

Analysis of the detector response function \cite{Jorma:2008aa,Jorma:2006aa,satz:2007aa} and a number of investigations of entanglement harvesting and quantum communication with (linear) UDW detectors \cite{Valentini:1991aa,Reznik:2003aa,Reznik:2005aa,Pozas:2015aa,Jonnson:2015aa,Blasco:2015aa,Edu:2015ab,Hotta:2009aa,Pozas:2016aa,Salton:2015aa} indicate that all   leading order UV divergences present in the time evolution of linearly coupled UDW detectors are regularizable.  While this is not the case for quadratically coupled detectors \cite{Iyer1980,Hinton,Takagi:1985aa,Takagi:1986aa},  it has been shown that all persistent divergences can also be renormalized for an individual quadratically coupled detector \cite{hummer}.  We will demonstrate below that a straightforward application of the leading-order prescription in \cite{hummer}  cannot renormalize persistent leading-order divergences in more complex scenarios with several detectors.  

\subsection{Time evolution of detector pairs}

Previous studies of the quadratic UDW model focused on the response of a single detector \cite{Iyer1980,Hinton,Takagi:1985aa,Takagi:1986aa,hummer}. Since one of our goals is to analyze the model dependence of vacuum entanglement harvesting, we will also consider the evolution of two particle detectors coupled to the field vacuum.

Both the linear and quadratically coupled UDW Hamiltonians can be rewritten for the two-detector case as
\begin{align} 
    \op H_{\op \phi}  &=\!\!\! \!\!\!\sum_{\nu\in\{\text{A},\text{B}\}}\!\!\!\!\!
    \lambda_{\nu} \chi_{\nu}(t-t_{\nu}) \op\mu_{\nu}(t) \! 
    \int  \! \d^n \bm{x} \, F_\nu(\bm{x}-\bm{x}_{\nu})  \op{\phi}(\bm{x},t),\label{hamlin2}\\
    \op H_{\op \phi^2} &=\!\!\! \!\!\!\sum_{\nu\in\{\text{A},\text{B}\}}\!\!\!\!\!
    \lambda_{\nu} \chi_{\nu}(t-t_{\nu}) \op\mu_{\nu}(t)\! \! 
    \int \! \d^n \bm{x}\,  F_\nu(\bm{x}-\bm{x}_{\nu}) \! :\!\op{\phi}^2(\bm{x},t)\!:,\!\label{hamquad2}
\end{align}
where $\nu\in \{\text{A},\text{B}\}$ is the label identifying  detectors A and B. Note that the coupling strength in the quadratic case does not have the same dimensions as in the linear case.    


If we let the initial state of the field-detector system be $\op \rho_{0}$, its time evolved state  is $\op \rho_{_T}=\op U \op \rho_{0} \op U^\dagger$, where the label $T$ denotes the timescale where the switching function is non-zero, and  the time evolution operator $\op U$ is given by the time-ordered exponential
\begin{equation}
\op U=\mathcal{T}\exp\left(\int_{-\infty}^\infty\!\!\!\!\!\d t\,\op H_\text{I}(t) \right).\label{timeordered}
\end{equation}

Consequently, we can express the time evolution operator $\op U$ in terms of a Dyson expansion as
\begin{align} 
	\op U = \openone +\op U^{(1)}+\op U^{(2)}+\mathcal{O}(\lambda^3_\mu),\label{expansion}
\end{align}
where
\begin{align} 
	\op U^{(1)} &= -\ii  \! \int_{-\infty}^\infty \!\!\!\! \d t  \, \op H_\text{I}(t)     \label{u1}      \\
	\op U^{(2)} &=  - \! \int_{-\infty}^{\infty} \!\!\!\!  \d t    \int_{-\infty}^{t} \!\!\!\!\!   \d t'  \, \op H_\text{I}(t)\op H_\text{I}(t') \label{u2}.
\end{align}
We can express $\op \rho_{_T}$ in a perturbative expansion as
\begin{align} 
	\op \rho_{_T}=\op \rho_0 \! + \! \op \rho^{(1,0)}_{_T} \! + \! \op \rho^{(0,1)}_{_T} \! + \! \op \rho^{(1,1)}_{_T} \! + \! \op \rho^{(2,0)}_{_T} \! + \! \op \rho^{(0,2)}_{_T} \! +\mathcal{O}(\lambda^3_\mu),\label{rho1}
\end{align}
where $\op \rho^{(i,j)}_{_T}=\op U^{(i)} \op \rho_{0} \op U^{(j)\dagger}$.

For our purposes we take as the initial state 
\begin{align} 
	\op \rho_{0}=\proj{0}{0}\otimes\op \rho_{\textsc{ab,0}}.\label{rho0}
\end{align}
with  the field starting out in its lowest-energy (vacuum) state. 

After time evolution, the time evolved partial state of the detectors is obtained by tracing out the field degrees of freedom:
\begin{equation} 
\op \rho_{{\aa\bb},\textsc{t}}=\tr_{\op\phi} \left(\op \rho_{_T}\right)
\end{equation}
The first order term $\op \rho^{(1,0)}_{_T} + \op \rho^{(0,1)}_{_T}$ does not contribute at all to the detectors' dynamics  for field states whose one-point function is zero. This includes Fock states, free thermal states and the vacuum state as a particular case of these two categories. In fact, for the vacuum state it can be easily proved that $\tr_{\op\phi} \left(\op \rho^{(i,j)}_{_T}\right)=0$ when $i+j$ is odd (see e.g., \cite{Pozas:2015aa}). Thus, we can express the time-evolved density matrix of the subsystem consisting of the two detectors as
\begin{align} 
	\op \rho_{{\aa\bb},\textsc{t}} 
	= \op \rho_\textsc{ab,0} 
	+ \lambda_\aa^2\op \rho_{\aa,\textsc{t}} 
	  +  \lambda_\bb^2\op \rho_{\bb,\textsc{t}} 
	+ \lambda_\aa\lambda_\bb\op \rho_{\text{cor},\textsc{t}}
	+ \mathcal{O}(\lambda^4_\mu),\label{detectorsstate}
\end{align}
where we have separated the local contributions to time evolution (proportional to $\lambda^2_{\aa}$ and $\lambda^2_{\bb}$ at leading order) from the non-local terms (responsible for the correlations the detectors acquire through the field) proportional to $\lambda_\aa\lambda_\bb$. Notice that, from \eqref{detectorsstate}, we can quickly recover the case of the evolution of a single detector just by taking $\lambda_\bb=0$.

Let us now particularize for the case where both detectors start out in the ground state:
\begin{align} 
	\op \rho_{{\aa\bb},0}=\proj{g_\aa}{g_\aa}\otimes\proj{g_\bb}{g_\bb}.\label{initialdetectorsstate}
\end{align}
It is convenient to pick the usual \cite{Pozas:2015aa} $4\times 4$ matrix representation for $\op\rho_{{\aa\bb},\textsc{t}}$ in the basis
\begin{align} 
\nonumber&\ket{g_\aa g_\bb}=(1,0,0,0)^\dagger,\quad &&\ket{e_\aa g_\bb}=(0,1,0,0)^\dagger,\\
&\ket{g_\aa e_\bb}=(0,0,1,0 )^\dagger,
\quad &&\ket{e_\aa e_\bb}=(0,0,0,1)^\dagger.
\label{basis}
\end{align}
In this basis, $\op\rho_{{\aa\bb},\textsc{t}}$ takes the form
\begin{align} 
	\op \rho_{{\aa\bb},\textsc{t}} = \! 
		\begin{pmatrix}
			 1\!-\!\mathcal{L}_{\aa\aa}\!\!-\!\mathcal{L}_{\bb\bb}\!\!\! & 0 & 0 & \mathcal{M} ^* \\
			 0 & \!\mathcal{L}_{\aa\aa} &  \mathcal{L} _{\aa\bb}\! & 0 \\
			 0 & \!\mathcal{L} _{\bb\aa} &  \mathcal{L}_{\bb\bb}\!  & 0 \\
			 \mathcal{M}  & 0 & 0 & 0 \\
		\end{pmatrix} \! \!
		+\!\mathcal{O}(\lambda^4_\mu)\label{densitymatrix},
\end{align}
where $\mathcal{M}$ and $\mathcal{L}_{\mu\nu}$ depend on the nature of the coupling (e.g., linear versus quadratic, different switching and smearing functions, etc. See sections \ref{LinCupSect} and \ref{QuadCupSect}). 


\subsubsection{Linear coupling \label{LinCupSect}}

The matrix elements of \eqref{densitymatrix} for the linear coupling have been studied at length in the literature (See, for instance,  \cite{Pozas:2015aa}, which sets the notation that we will follow here) and are given by
\begin{align} 
	\mathcal{M}^{\op\phi} \! =	&-\lambda_\aa\lambda_\bb
		\int_{-\infty}^{\infty} \!\!\!\!\! \d t  
		\int_{-\infty}^{t} \!\!\!\!\! \d t' \!
		\int \d^n \bm{x}  
		\int  \d^n \bm{x}' \, 
		\notag\\ & \times\label{linearM}
		M(t ,\bm{x} ,t'\!,\bm{x}') \,
		 W^{\op\phi}
		(t,\bm{x},t'\!,\bm{x}')\\[3mm]
	 \mathcal{L} _{\nu\mu}^{\op\phi} \! = & \lambda_\nu\lambda_\mu
		\int_{-\infty}^{\infty} \!\!\!\!\! \d t  
		\int_{-\infty}^{\infty} \!\!\!\!\! \d t' \!
		\int  \d^n \bm{x}  
		\int  \d^n \bm{x}' \,
		\label{linearL}\\ & \times\notag
		L^*_\nu (t ,\bm{x} ) \,
		 L_\mu (t'\!,\bm{x}') \, 
		W^{\op\phi} (t,\bm{x},t'\!,\bm{x}'),
\end{align}
where, assuming $\chi$ and $F$ are real,  $	L_{\nu}$ and $M(t,\bm{x},t'\!,\bm{x}')$ are given by
\begin{align} 
	L_{\nu} (t,\bm{x})=& \,
		\chi_{\nu}(t-t_\nu) \, F_{\nu}(\bm{x}-\bm{x}_\nu)  \, e^{\ii \Omega_{\nu} t}\label{linearlittleL}\\
    M(t,\bm{x},t'\!,\bm{x}')=&\,
        L_{\aa} (t,\bm{x})
        L_{\bb} (t'\!,\bm{x}')
        \notag\\&
        \!+\!
        L_{\aa} (t'\!,\bm{x}')
        L_{\bb} (t,\bm{x}),\!\label{linearlittleM}
\end{align}
 and the Wightman function, $W^{\op\phi}$, is given by
\begin{align} 
	W^{\op\phi} (t,\bm{x},t'\!,\bm{x}')=& \, 
		\bra{0}  \op\phi(\bm{x},t) \, \op\phi(\bm{x}'\!,t')  \ket{0}.\label{twopointLIN}
\end{align}

To find an explicit expression for the Wightman function, we will utilize a plane-wave mode expansion of the field operator with soft UV cutoff $\epsilon$,
\begin{align}  
	\op\phi (\bm{x},t) \!& = \!\! 
		\int  \frac{\d^n \bm{k} \, 
		e^{-\epsilon|\bm{k}|/2}}{\sqrt{2(2\pi)^n|\bm{k}|}} \! 
		\notag\\&\times
		\left(  \! 
		    \hat{a}^\dagger_k e^{\ii (|\bm{k}|t-\bm{k}\cdot\bm{x})} \! + 
		    \hat{a}^{\phantom{\dagger}}_k e^{-\ii (|\bm{k}|t-\bm{k}\cdot\bm{x})}  \! \right) \!.\label{planewave}
\end{align}
 Here, $\hat{a}^\dagger_{\bm{k}}$ (and $\hat{a}_{\bm{k}}$) are creation (and annihilation) operators which satisfy the canonical  commutation relations
\mbox{$[\hat{a}_{\bm{k}_1}^{\phantom{\dagger}},\hat{a}_{\bm{k}_2}^\dagger ]=\delta^{(n)}(\bm{k}_1-\bm{k}_2)$}.

Typically, the introduction of $\epsilon$ could be associated with a regularization procedure that leads to the usual pole prescription, in which the limit $\epsilon\rightarrow0$ is well-defined and eventually taken when evaluating observable quantities. However $\epsilon$ can also be viewed as an \textit{ad hoc} screening of the detector's sensitivity to high frequency modes of the field (soft UV cutoff). This would effectively model, for example, a frequency dependent coupling strength where a detector  does not couple to frequencies much larger than $\epsilon^{-1}$. When giving this kind of interpretation to the $\epsilon $-regularization one should be careful with possible non-localities introduced in the theory due to a finite value of $\epsilon$ \cite{Edu:2015aa}. Although this point will not be relevant when the limit $\epsilon\rightarrow0$ is well defined, it  must be taken into account when managing possibly UV divergent terms, especially in the case of the quadratic coupling \eqref{hamquad}, as we will see below.

The Wightman function \eqref{twopointLIN} for the linear coupling case can be written as
\begin{align} 
        W^{\op\phi} (t,\bm{x},t'\!,\bm{x}') 		= 
		\! \int \! \d^n \bm{k}\,
		\frac{e^{\ii\left(|\bm{k}|(t'\!-t)-\bm{k}\cdot(\bm{x}'\!-\bm{x})\right)-|\bm{k}|\epsilon}}{2(2\pi)^{n} |\bm{k}|} \label{twopointLIN2},
\end{align}
as it is easy to check, for example, through the usual plane-wave expansion in Eq.~\eqref{planewave}. Particularizing to 3+1 dimensions, the two-point function becomes
\begin{align} 
	W^{\op\phi} (t,\bm{x},t',\bm{x}') 		= \frac{1}{4\pi^2(\bm{x}-\bm{x'})^2-(t-t'-\ii\epsilon)^2}. \label{twopointLIN3+1}
\end{align}
Note here that we see $\epsilon$ takes the form of the usual pole prescription for the Wightman function. 

\subsubsection{Quadratic coupling\label{QuadCupSect}}

For the quadratic coupling case in \eqref{hamquad2}, the elements of the density matrix \eqref{densitymatrix} take the following form  
\begin{align} 
	\mathcal{M}^{\op\phi^2} \! =	&-\lambda_\aa\lambda_\bb
		\int_{-\infty}^{\infty} \!\!\!\!\! \d t  
		\int_{-\infty}^{t} \!\!\!\!\! \d t' \!
		\int \d^n \bm{x}  
		\int  \d^n \bm{x}' \, 
		\notag\\ & \times
		M(t ,\bm{x} ,t'\!,\bm{x}') \, W^{\op\phi^2}
		(t,\bm{x},t'\!,\bm{x}')\label{quadM}\\
	 \mathcal{L} ^{\op\phi^2}_{\nu\mu} \! = & \lambda_\nu\lambda_\mu
		\int_{-\infty}^{\infty} \!\!\!\!\! \d t  
		\int_{-\infty}^{\infty} \!\!\!\!\! \d t' \!
		\int  \d^n \bm{x}  
		\int  \d^n \bm{x}' \,
		\notag\\ & \times
		L^*_\nu (t ,\bm{x} ) \, L_\mu (t'\!,\bm{x}') \, 
		W^{\op\phi^2} (t,\bm{x},t'\!,\bm{x}'),\label{quadL}
\end{align}
which is structurally the same as in the linear coupling case. Indeed,  $M$ and $L_\nu$ are also defined as in Eqs. \eqref{linearlittleM} and \eqref{linearlittleL}, respectively. Thus, the difference between the usual UDW detector and the quadratically coupled UDW detector comes at the level of the functional $W^{\op\phi^2}$. For the quadratically coupled model, $W^{\op\phi^2}$ is the vacuum expectation of the \textit{normal ordering} of the square of the field operator at two different points, as given by
\begin{align} 
	W^{\op\phi^2} (t,\bm{x},t'\!,\bm{x}')=& \, 
		\bra{0}  \!:\! \op\phi^2(\bm{x},t) \! : \, : \! \op\phi^2(\bm{x}'\!,t') \! :\! \ket{0}.\label{twopointQUAD}
\end{align}

In Appendix \ref{ap:wick} we show that the correlation functions $W^{\op\phi}$ and $W^{\op\phi^2}$ satisfy the following relation
\begin{align}  
	W^{\op\phi^2} (t,\bm{x},t'\!,\bm{x}') 	\!	= 2W^{\op\phi} (t,\bm{x},t'\!,\bm{x}') ^2 \label{WickTrick},
\end{align}
which allows us to write $W^{\op\phi^2}$ explicitly as
\begin{align}  
		W&^{\op\phi^2} \! (t,\bm{x},t'\!,\bm{x}')	= 
		\!\! \int \!\! \d^n  \bm{k}_1
		\!\! \int \!\! \d^n  \bm{k}_2 
		\frac{  (2\pi)^{-2n}  }
		{ |\bm{k}_1| |\bm{k}_2| }\!\label{twopointQUAD3+1}\\
		\times& e^{\ii\left(\left(|\bm{k}_2|+|\bm{k}_1|\right)\left(t'-t\right)-\left(\bm{k}_2+\bm{k}_1\right)\cdot\left(\bm{x}'-\bm{x}\right)\right)-\left(|\bm{k}_1|+|\bm{k}_2|\right)\epsilon}\notag.
\end{align}
If we particularize to 3+1 dimensions, the correlation function $W^{\op\phi^2}$ is
\begin{align}  
	W^{\op\phi^2}\! (t,\bm{x},t'\!,\bm{x}') 	\!	= \frac{2}{\big[4\pi^{2}(\bm{x}-\bm{x'})^2\!-(t-t'\!-\ii\epsilon)^2\big]^2} \label{twopointQUAD2}.
\end{align}

Note here that the correlator for the quadratic UDW detector has a higher power polynomial in $\bm x$ and $t$ in its denominator than does the usual correlator for the linear UDW detector.

\section{Single detector vacuum excitation probability in 3+1 dimensions, $\mathcal{L}_{_{\text{A}\text{A}}}$ 
\label{singledetector}}

\begin{table}[ht]
\begin{ruledtabular}
{\setlength{\extrarowheight}{5pt}\begin{tabular}{ccl}
Dimensionless\\variable & Expression & Physical meaning\\[5pt]
\hline & &\\[-13pt]
\hline$\alpha$ & $\Omega T$  & Energy gap\\[5pt]
\hline$\eta$ & $\epsilon T$  & UV cutoff\\[5pt]
\hline$\beta$ & $|\bm x_\aa-\bm x_\bb|/T$ & Detectors' separation\\[5pt]
\hline$\gamma_\nu $ & $t_\nu T$ & switch-on times \\[5pt]
\hline$\delta$ & $\sigma /T$ & Detectors' size\\[5pt]
\hline$\xi$   & $q/ T$ & ---\\[3pt]
\end{tabular}}
\caption{Collection of all the dimensionless quantities that are used throughout this paper.}\label{tab:parametersTable}
\end{ruledtabular}
\end{table}

In this section we will briefly review the vacuum excitation probability (VEP) of a single detector for the usual linear UDW detector model and the more recently studied VEP for a quadratically coupled model \cite{hummer}. The vacuum excitation probability is the probability of excitation of a single UDW detector initialized in its ground state in the vacuum. 

It is well known that in 3+1 dimensions, point-like linearly coupled UDW detectors with sharp switching functions suffer regularizable UV divergences, which we recall can be eliminated by introducing a smooth switching or smearing function \cite{Jorma:2006aa,satz:2007aa,Jorma:2008aa}. However, quadratic UDW models (such as the quadratic scalar model introduced by Hinton in \cite{Hinton}, the cavity detector coupled to a fermionic field \cite{Iyer1980}, or the fermionic UDW-like detector model introduced in \cite{Takagi:1985aa,Takagi:1986aa}) have VEPs that present persistent divergences (not removable with a smooth switching and/or smearing). These persistent divergences can, however, be renormalized with techniques analogous to those in QED \cite{hummer}. Once renormalized, the quadratically coupled single-detector UDW model is regularizable both in its scalar and fermionic variants.

To find the time evolved state of a single (quadratically or linearly coupled) detector, we begin with the density matrix \eqref{densitymatrix}, then set $\lambda_\bb=0$. It is then simple to trace out detector B to find the single-detector reduced state,
\begin{align}  
	\op \rho_{\aa,_T} =\mathrm{Tr}_{\bb}\left(\op \rho_{{\aa\bb},_T}\right) =
		\begin{pmatrix}
			 1- \mathcal{L}_{{\aa\aa}}\!\!  & 0 \\
			 0 &  \!\!\mathcal{L}_{{\aa\aa}}   \\
		\end{pmatrix}
		\!+\!\mathcal{O}(\lambda^4_\aa),\label{singledensitymatrix}
\end{align}
in the basis $\ket{g_\aa}=(1,0)^\dagger$, $\ket{e_\aa}=(0,1)^\dagger$.
The element of \eqref{singledensitymatrix}, $\mathcal{L}_{{\aa\aa}}$, given in Eq.~\eqref{linearL}, is the vacuum excitation probability of detector A.

The vacuum excitation probability expressed as Eq.~\eqref{linearL} is quite general and can be particularized to any spacetime dimensionality, switching function and spatial profile. For this analysis we will use smooth switching and spatial smearing functions which are only strongly supported in a finite region ($T$ and $\sigma$ respectively). Smooth switching and smearing will ensure the removal of all regularizable divergences of the kind studied in \cite{Jorma:2008aa}. In particular, we choose Gaussian switching and Gaussian smearing, 
\begin{align}  
	F_\nu(\bm{x}-\bm{x}_\nu)=\frac{1}{(\sqrt{\pi}\sigma)^n}e^{-(\bm{x}-\bm{x}_\nu)^2/\sigma^2},\label{spatprof}
\end{align}
\begin{align}  
		\chi_\nu(t-t_\nu)=e^{-(t-t_\nu)^2/T^2}.\label{switchingFunct}
\end{align}

As mentioned previously, in the literature UDW detectors are often considered to be point-like. Notice that the Gaussian spatial profile can be particularized to the point-like case by taking the limit $\sigma\rightarrow0$.

\subsection{Linear coupling, $\mathcal{L}^{\op \phi}_{_{\text{A}\text{A}}}$}
In this section we will calculate the vacuum excitation probability for a single linearly-coupled, 3+1 dimensional UDW detector with Gaussian switching and smearing functions. To do so, we first begin with Eq.~\eqref{linearL}, setting $\mu=\nu=\text{A}$. Then we substitute into Eq.~\eqref{linearL} the 3+1 dimensional Wightman function  \eqref{twopointLIN3+1}, the spatial profile \eqref{spatprof}, and the switching function \eqref{switchingFunct}. 
Furthermore, to further simplify the calculation, we apply the  change of coordinates
\begin{align}    
u=& \, t_1+t_2, &
v=& \, t_1-t_2, \notag\\
\bm{p}=& \, \bm{x}_1+\bm{x}_2, &
\bm{q}=& \, \bm{x}_1-\bm{x}_2 \label{coords1}.
\end{align}
This results in
\begin{align}  
	\mathcal{L}^{\op \phi}_{\aa\aa} = &
	\frac{\lambda ^2  }{64 \pi ^5 \sigma ^6  }
	\int_{-\infty}^{\infty}\!\!\!\d u\,
	e^{	        -\frac{u^2}{2 T^2}  }
	\int\!\d^3 \bm p\,
	e^{	        -\frac{\bm p^2}{2 \sigma ^2}}
	\notag\\&\times
	\int_{-\infty}^{\infty}\!\!\!\d v
    \int\!\d^3 \bm q\,
    \frac{
	    e^{
	        -\frac{\bm q^2}{2 \sigma ^2}
	        -\frac{v^2}{2 T^2} 
	        -\ii  v \Omega }
	}{\left(\bm q^2-(v -\ii  \epsilon )^2\right)}
	.\label{VacExitProb22}
\end{align}
The above integrals in  $u$, $\bm p$ and the angular parts of  $\bm q$ can be easily evaluated in closed form. To find the integral over $v$, we use the convolution theorem, as outlined in the Appendix \ref{ap:convolution}. At this point, it is convenient to follow \cite{Pozas:2015aa} and rewrite these intergals in terms of dimensionless parameters $\alpha,\beta,\gamma,\delta,\eta,$ and $\xi$ as outlined in Table \ref{tab:parametersTable}. The outcome is
\begin{align}  
    \mathcal{L}&^{\op \phi}_{\aa\aa} \!=
    -\frac{\lambda ^2\ii }{8 \pi  \delta ^3 }
    \int_0^\infty \!\!\! \d \xi \,
    \xi  e^{\alpha  \eta  -\ii  \alpha  \xi -\frac{\xi ^2}{2 \delta ^2}+\frac{\eta ^2}{2} -\ii  \eta  \xi -\frac{\xi ^2}{2}} 
       \label{Lfin24} \\&\times\notag\!
    \Bigg[\!
        \,\text{erfc}\left(\frac{\alpha +\eta -\ii \xi }{\sqrt{2}}\right)-e^{2 \ii \xi  (\alpha +\eta )} \,\text{erfc}\left(\frac{\alpha +\eta +\ii \xi }{\sqrt{2}}\right)\!
    \Bigg]
    ,
\end{align}
where erfc is the complementary error function, defined in terms of the error function as follows:
\begin{align}
    \,\text{erf}\left(z\right)&=
    \frac{2}{\sqrt{\pi }}\int _0^zd t e^{-t^2}\label{errorfunct}\\
    \,\text{erfc}\left(z\right)&=
    1 -\,\text{erf}\left(z\right)\label{comperrorfunct}.
\end{align}

At this point, we can take the UV cutoff scale to infinity ($\epsilon\rightarrow0$, or in dimensioless quantities, \mbox{$\eta\to0$}, as per table \ref{tab:parametersTable}). The result is
\begin{align} 
    \,\lim_{\eta\to0}&\mathcal{L}^{\op \phi}_{\aa\aa} \!=
    \frac {-\ii  \lambda ^2 }{8 \pi  \delta ^3}
    \int_0^\infty \!\!\! \d \xi \,
    \xi  e^{-\frac{\xi  \left( 2 \ii \alpha  \delta ^2+\delta ^2 \xi +\xi \right)}{2 \delta ^2}} 
    \notag\\&\!\!\!\!\!\!\times \!
    \Bigg[
        \,\text{erfc}\left(\frac{\alpha  -\ii  \xi }{\sqrt{2}}\right)
        -e^{ 2 \ii \alpha  \xi } \,\text{erfc}\left(\frac{\alpha  +\ii  \xi }{\sqrt{2}}\right)
    \Bigg]
    \label{Lfin37}
    ,
\end{align}
which is not divergent.  Fig.~\ref{fig:allconvergencePlots}.(a) illustrates the behaviour of $\mathcal{L}^{\op \phi}_{\aa\aa}$ as $\eta\to0$ is reached.

We note that in previous literature closed expressionss for \eqref{Lfin37} have been found for Gaussian switching and smearing functions in 3+1 dimensions \cite{Pozas:2015aa}. The difference between calculations here and in \cite{Pozas:2015aa} is that we have worked in the position representation. One can readily check numerically that all elements $\mathcal{L}_{\mu\nu}$ and $\mathcal{M}$ in this paper are equivalent (after the limit $\eta\to0$ is taken) to those in \cite{Pozas:2015aa} for the linear detector. The motivation behind complicating the calculation of the linear matrix elements by working in the position representation lies in the difficulty of calculating the matrix elements of the quadratic detector pairs, which is reduced by the method described here. Moreover, there is an additional advantage working in the position representation in the linear case: the method of computing leading order density matrix elements in the position representation used here yields results that have greater numerical stability for small detector gap in those terms for which we do not have closed expressions neither in position nor in momentum representation in \cite{Pozas:2015aa}, as we will show when we present numerical results in section \ref{harvesting}.

\subsection{Quadratic coupling, $\mathcal{L}^{\op \phi^2}_{_{\text{A}\text{A}}}$\label{quadVEP}}

Similar to the linear model, in this section we calculate the vacuum excitation probability for the quadratic model, $\mathcal{L}^{\op \phi}_{\aa\aa}$. We begin with Eq.~\eqref{linearL}, setting $\mu=\nu=\text{A}$. Then we substitute into \eqref{linearL} the quadratic two-point correlator in  3+1 dimensions \eqref{twopointQUAD2}, the spatial profile \eqref{spatprof}, and the switching function \eqref{switchingFunct}.  Applying the same change of coordinates as in the linear case, \eqref{coords1}, yields
\begin{align}  
    \mathcal{L}_{\aa\aa}^{\op \phi^2}=&
    \frac{\lambda ^2}{128 \pi ^7 \sigma ^6 }
	\int_{-\infty}^{\infty}\!\!\!\!\d u\,
	e^{-\frac{u^2}{2 T^2}}
	\int\!\d^3 \bm p\,
	e^{-\frac{\bm p^2}{2 \sigma ^2}}
	\notag\\&\times
	\int_{-\infty}^{\infty}\!\!\!\!\d v
	\int\!\d^3 \bm q
	\frac{
	e^{
	-\frac{\bm q^2}{2 \sigma ^2}
	-\frac{v^2}{2 T^2}
	 -\ii  v \Omega 
	}}{\big(\bm q^2-(v -\ii  \epsilon )^2\big)^2}
    \label{quadVEP1}.
\end{align}
The integrals over $u$, $\bm p$, the angular part of $\bm q$, and  $v$ can be evaluated in closed form (again, the last performed through a convolution product as shown in appendix \ref{ap:convolution}). Once again, it is convenient to utilize the convention in \cite{Pozas:2015aa} and recast these integrals in terms of the dimensionless parameters $\alpha,\beta,\gamma,\delta,\eta,$ and $\xi$ (outlined in Table \ref{tab:parametersTable}).  The result is
\begin{align}  
    \mathcal{L}&_{\aa\aa}^{\op \phi^2}=
    \frac{
        \lambda ^2
        e^{\alpha  \eta
        +\frac{\eta ^2}{2}}
        }{32 \pi ^4 \delta ^3  T^3}
    \int_0^\infty\d q
    \frac{e^{ -\ii  \alpha  \xi 
    -\frac{\xi ^2}{2 \delta ^2}
     -\ii  \eta  \xi 
    -\frac{\xi ^2}{2}} }{\xi }
	\notag\\&\times
    \Bigg[
        -2 \sqrt{2 \pi } \xi  e^{-\frac{1}{2} (\alpha +\eta  -\ii  \xi )^2}
    	+\pi  \bigg[
	    \alpha  \xi+\eta  \xi  -\ii  (\xi ^2 +1)  
    	\notag\\&\quad\quad
	    +e^{ 2 \ii \xi  (\alpha +\eta )} \left(\alpha  \xi +\eta  \xi  +\ii  \left(\xi ^2+1\right)\right)
    	\notag\\&\quad\quad\times
	    \,\text{erfc}\left(\frac{\alpha +\eta  +\ii  \xi }{\sqrt{2}}\right)
    	\bigg]
	    \notag\\&\quad
    	+\pi  \left(\ii \alpha  \xi  +\ii  \eta  \xi +\xi ^2+1\right)     \,\text{erfi}\left(\frac{\ii \alpha  +\ii  \eta +\xi }{\sqrt{2}}\right)
    \Bigg]
    \label{quadVEP2},
\end{align}
where erfi is the imaginary error function defined as
\begin{align}
    \,\text{erfi}(z)=-\ii\,\,\text{erf} (\ii z)\label{imagerrorfunct}.
\end{align} 

For this integrand, the limit of no cutoff, i.e, \mbox{$\frac{\epsilon}{T}=\eta\rightarrow0$}, at constant $T$, is well-defined:
\begin{align}  
     \lim_{\epsilon\to0}\mathcal{L}&_{\aa\aa}^{\op \phi^2}=
    \frac{\lambda ^2}{32 \pi ^4 \delta ^3  T^2}
    \int_0^\infty
    \frac{\d \xi}{\xi }
    e^{-\frac{\alpha ^2}{2} -\ii  \alpha  \xi -\frac{1}{2} \left(\frac{1}{\delta ^2}+1\right) \xi ^2} 
    \notag\\&\times
    \Bigg[
         -\ii  \pi  e^{\frac{\alpha ^2}{2}} \xi ^2
        +\pi  e^{\frac{\alpha ^2}{2}} \alpha  \xi 
         -\ii  \pi  e^{\frac{\alpha ^2}{2}}
        \notag\\&
         +\ii  \pi  \xi ^2 e^{\frac{1}{2} \alpha  (\alpha +4 \ii \xi )}
        +\pi  \alpha  \xi  e^{\frac{1}{2} \alpha  (\alpha +4 \ii \xi )}
        \notag\\&
        -2 \sqrt{2 \pi } \xi  e^{\frac{1}{2} \xi  (\xi + 2 \ii \alpha )}
         +\ii  \pi  e^{\frac{1}{2} \alpha  (\alpha +4 \ii \xi )}
        \notag\\&
        -\pi  e^{\frac{\alpha ^2}{2}} \Big(\alpha  \xi  -\ii  \left(\xi ^2+1\right)\Big) \,\text{erf}\left(\frac{\alpha  -\ii  \xi }{\sqrt{2}}\right)
        \notag\\&
        -\pi  e^{\frac{1}{2} \alpha  (\alpha +4 \ii \xi )} \Big(\alpha  \xi  +\ii  \left(\xi ^2+1\right)\Big) \,\text{erf}\left(\frac{\alpha  +\ii  \xi }{\sqrt{2}}\right)
    \Bigg]
    \label{quadVEP3}.
\end{align}
This integral is convergent. How the convergent $\eta\to 0$ limit is reached is shown numerically in Fig.~\ref{fig:allconvergencePlots}.(d).

\section{The two-detector model 
\label{twodetector}}

The vacuum excitation probability does not provide full information about the time evolution of a pair of particle detectors coupled to the field. Indeed, to characterize more complicated effects, such as entanglement harvesting \cite{Reznik:2003aa,Valentini:1991aa,Pozas:2015aa}, or quantum communication \cite{Blasco:2015aa,Blasco:2016aa,Jonnson:2015aa,Jonsson:2014aa,Jonnson:2015ab,Edu:2015aa,Landulfo}, the full time-evolved density matrix of two detectors coupled to the field is necessary. The detectors' time-evolved density matrix \eqref{densitymatrix} has extra terms in addition to the VEPs. Two different kinds of non-local  terms, $\mathcal{L}_{\aa\bb}$ and $\mathcal{M}$, now appear along with their complex conjugates. To fully characterize the two detector system, we need to find explicit expressions for these terms and study the regularity of their behaviour.

As we will discuss in detail below (and as mentioned in \cite{Pozas:2015aa}),  $\mathcal{L}_{\aa\bb}$ is the term responsible for the leading order contribution to classical correlations (or, possibly, discord)  between the detectors, whereas $\mathcal{M}$ can be thought of as responsible for the harvested entanglement from the field to the detectors, as we will discuss in section \ref{entharvesting}. We will analyze these two terms independently in the next two subsections.

\subsection{$\mathcal{L}_{_{\text{A}\text{B}}}$ non-local term in 3+1 dimensions}

In the following, we will find  $\mathcal{L}_{\aa\bb}$ for the linear and quadratic models in 3+1 dimensions.a

\subsubsection{Linear coupling, $\mathcal{L}^{\op\phi}_{\textsc{a}\textsc{b}}$}
For the linear UDW detector, the term $\mathcal{L}^{\op\phi}_{\textsc{a}\textsc{b}}$ is given by Eq.~\eqref{linearL} when $\nu=\text{A}$ and $\mu=\text{B}$. We also explicitly write the Wightman function in 3+1 dimensions \eqref{twopointLIN3+1}, the spatial profile \eqref{spatprof}, and the switching function \eqref{switchingFunct} in equation \eqref{linearL}.  The same change of coordinates as in the calculation of $\mathcal{L}_{\aa\aa}$, shown in Eq.~\eqref{coords1}, again simplifies the calculation. This transformation yields
\begin{align}  
	\mathcal{L}^{\op \phi}_{\aa\bb} = &
	\frac{\lambda ^2e^{-\frac{ t_{\aa} ^2}{T^2}-\frac{ t_{\bb} ^2}{T^2}-\frac{ \bm x_{\aa} ^2}{\sigma ^2}-\frac{ \bm x_{\bb} ^2}{\sigma ^2}}}{64 \pi ^5 \sigma ^6 }
	\int_{-\infty}^{\infty}\!\!\!\!\d u\,
	e^{
        +\frac{ t_{\aa}  u}{T^2}
        +\frac{ t_{\bb}  u}{T^2}
        -\frac{u^2}{2 T^2}
	}
	\notag\\&\times
	\int\!\d^3 \bm p
	e^{
            -\frac{\bm p^2}{2 \sigma ^2}
            +\frac{\bm p  \bm x_{\aa} }{\sigma ^2}
            +\frac{\bm p  \bm x_{\bb} }{\sigma ^2}
	}	
	\int_{-\infty}^{\infty}\!\!\!\d v
	\int\!\d^3 \bm q
	\notag\\&\times
	\frac{
        e^{
            -\frac{\bm q^2}{2 \sigma ^2}
            +\frac{\bm q  \bm x_{\aa} }{\sigma ^2}
            -\frac{\bm q  \bm x_{\bb} }{\sigma ^2}
            +\frac{ t_{\aa}  v}{T^2}
            -\frac{ t_{\bb}  v}{T^2}
            -\frac{v^2}{2 T^2}
            -\ii v \Omega }}
	{\bm q^2-(v-\ii \epsilon )^2}
	.\label{VacExitProb23}
\end{align}
The integrals over $u$, $\bm p$, the angular part of $\bm q$, and  $v$ can be evaluated in closed form (with the same technology shown in Appendix \ref{ap:convolution}). As before, we follow \cite{Pozas:2015aa} and rewrite these integrals in terms of the dimensionless parameters as outlined in Table \ref{tab:parametersTable}.  The result is
\begin{align}  
    \mathcal{L}^{\op \phi}_{\aa\bb} \!&=
     \frac{\lambda ^2    
     e^{
        -\frac{ \beta  ^2}{2 \delta ^2}
        -\frac{1}{2} ( \gamma_{\aa} - \gamma_{\bb} )^2
    }
     e^{
        \alpha  \eta 
        +\ii  \gamma_{\aa} \eta 
        -\ii  \gamma_{\bb} \eta 
        +\frac{\eta ^2}{2}
        }
    }{8 \pi  \delta    \beta   }
   \notag\\&\times
    \int_0^\infty\!\!\!\d \xi\,
   \sinh \left(\frac{\xi   \beta   }{\delta ^2}\right) 
   \bigg[
        \ii \left(e^{2 \xi  (\ii \alpha + \gamma_{\bb}+\ii  \eta )}-e^{2  \gamma_{\aa} \xi }\right)
   \notag\\&
        +e^{2 \xi  (\ii \alpha + \gamma_{\bb}+\ii  \eta )} \,\text{erfi}\left(\frac{-\ii \alpha + \gamma_{\aa}- \gamma_{\bb}-\ii \eta +\xi }{\sqrt{2}}\right)
   \notag\\&
        +e^{2  \gamma_{\aa} \xi } \,\text{erfi}\left(\frac{\ii \alpha - \gamma_{\aa}+ \gamma_{\bb}+\ii  \eta +\xi }{\sqrt{2}}\right)
    \bigg] 
   \notag\\&\times
   e^{
        -\ii \alpha  \xi 
        - \gamma_{\aa} \xi 
        - \gamma_{\bb} \xi 
        -\frac{\xi ^2}{2 \delta ^2}
        -\ii \eta  \xi 
        -\frac{\xi ^2}{2}}
    \label{notyetlabeled1},
\end{align}
which has a well-defined limit as $\eta\to0$,
\begin{align}  
    \lim_{\eta\to0}\mathcal{L}^{\op \phi}_{\aa\bb} \!&=
    \frac{\ii \lambda ^2 
    e^{
        -\frac{ \beta  ^2}{2 \delta ^2}
        -\frac{1}{2} ( \gamma_{\aa} - \gamma_{\bb} )^2
    }    }{8 \pi  \delta   \left|   \beta   \right| }
    \int_0^\infty\!\!\!\d \xi\,
    \notag\\&\times
    \sinh \left(\frac{\xi  \left|   \beta   \right| }{\delta ^2}\right) 
    e^{
        -\ii \alpha  \xi 
        - \gamma_{\aa}  \xi 
        - \gamma_{\bb}  \xi 
        -\frac{\xi ^2}{2 \delta ^2}
        -\frac{\xi ^2}{2}
    }
    \notag\\&\times
    \bigg[e^{2 \xi  ( \gamma_{\bb}  +\ii  \alpha )} \,\text{erfc}\left(\frac{\alpha  +\ii  ( \gamma_{\aa} - \gamma_{\bb} +\xi )}{\sqrt{2}}\right)
    \notag\\&\quad 
    -e^{2  \gamma_{\aa}  \xi } \,\text{erfc}\left(\frac{\alpha  +\ii  ( \gamma_{\aa} - \gamma_{\bb} -\xi )}{\sqrt{2}}\right)\bigg] 
    \label{notyetlabeled2}.
\end{align}
Not only is the integrand well-defined, but the integral is convergent as well. We show numerically how the convergent limit $\eta\rightarrow0$ of $|\mathcal{L}^{\op \phi}_{\aa\bb}|$ is reached in Fig.~\ref{fig:allconvergencePlots}.(b).

\subsubsection{Quadratic coupling, $\mathcal{L}^{\op\phi^2}_{{\textsc{a}\textsc{b}}}$}

In order to find $\mathcal{L}^{_{\op \phi^2}}_{\aa\bb}$, given by Eq.~\eqref{linearL}, we set $\nu=\text{A}$ and $\mu=\text{B}$. We then substitute into Eq.~\eqref{linearL} the quadratic two-point correlator in 3+1 dimensions \eqref{twopointQUAD2}, the spatial profile \eqref{spatprof}, and the switching function \eqref{switchingFunct}.  As is now tradition, we will do the same change of coordinates as in the calculation of the VEPs, shown in Eq.~\eqref{coords1}. This transformation results in
\begin{align}  
    \mathcal{L}&_{\aa\bb}^{\op \phi^2}=
    \frac{\lambda ^2 e^{-\frac{ t_{\aa} ^2+ t_{\bb} ^2}{T^2}-\frac{ \bm x_{\aa} ^2+ \bm x_{\bb} ^2}{\sigma ^2}}
    }{128 \pi ^7 \sigma ^6 }
	\notag\\&\times
	\int_{-\infty}^{\infty}\!\!\!\d u
	e^{
        \frac{ (t_{\aa}+t_{\bb} )  u}{T^2}
        -\frac{u^2}{2 T^2}}
	\int\!\d^3 \bm p\,
	e^{
        \frac{\bm p  (\bm x_{\aa} +\bm x_{\bb})}{\sigma ^2}
        -\frac{\bm p^2}{2 \sigma ^2}}
	\notag\\&\times
	\int_{-\infty}^{\infty}\!\!\!\d v\!\!
	\int\!\d^3 \bm q
    \frac{
    e^{
        \frac{\bm q  (\bm x_{\aa}-\bm x_{\bb}) }{\sigma ^2}
        +\frac{ v(t_{\aa} -t_{\bb} ) }{T^2}
        -\frac{\bm q^2}{2 \sigma ^2}
        -\frac{v^2}{2 T^2}
        -\ii v \Omega }
    }{ \left(\bm q^2-( v-\ii \epsilon )^2\right)^2}
    \label{quadVEP4}.
\end{align}
The integrals over $u$, $\bm p$, the angular part of $\bm q$, and  $v$ can be evaluated in closed form (details in appendix \ref{ap:convolution}). Once again, we follow \cite{Pozas:2015aa} and rewrite these integrals in terms of the dimensionless quantities outlined in Table \ref{tab:parametersTable}.  The result is
\begin{align}  
	\mathcal{L}&_{\aa\bb}^{\op \phi^2}  = 
	\frac{\lambda ^2 	e^{
	    \alpha  \eta 
	    -\frac{( \beta  )^2}{2 \delta ^2}
	    -\frac{1}{2} ( \gamma_{\aa}- \gamma_{\bb})^2
	    +\ii  \gamma_{\aa} \eta 
	    -\ii  \gamma_{\bb} \eta 
	    +\frac{\eta ^2}{2}
	}}{32 \pi ^4 \delta   T^2 ( \beta  )}
	\notag\\\times &
	\int_0^\infty\d \xi
	\frac{		e^{
	    -\ii \alpha  \xi 
	    + \gamma_{\aa} \xi 
	    - \gamma_{\bb} \xi 
	    -\frac{\xi ^2}{2 \delta ^2}
	    -\ii \eta  \xi 
	    -\frac{\xi ^2}{2}
	}}{\xi ^2}
	\notag\\\times&
	\sinh \left(\frac{\xi  ( \beta  )}{\delta ^2}\right) 
	\Bigg[
	    -2 \sqrt{2 \pi } \xi  e^{-\frac{1}{2} (\alpha +\ii  \gamma_{\aa}-\ii ( \gamma_{\bb}+\ii \eta +\xi ))^2}
	\notag\\&
	    +\pi  \left(\ii \alpha  \xi - \gamma_{\aa} \xi + \gamma_{\bb} \xi +\ii \eta  \xi +\xi ^2+1\right) 
	\notag\\& \times
	    \,\text{erfi}\left(\frac{\ii \alpha - \gamma_{\aa}+ \gamma_{\bb}+\ii \eta +\xi }{\sqrt{2}}\right)
	\notag\\&
	    +\pi  \bigg[
	        \alpha  \xi 
	        +\ii  \gamma_{\aa} \xi 
	        -\ii  \gamma_{\bb} \xi 
	        +\eta  \xi -\ii \xi ^2-\ii 
	\notag\\&\quad
	        +e^{2 \xi  (\ii \alpha - \gamma_{\aa}+ \gamma_{\bb}+\ii \eta )} 
	        \left(\xi\left(\gamma_{\aa}  - \gamma_{\bb}  -\ii \alpha    -\ii \eta   +\xi\right) +1\right)
	\notag\\&\quad\times 
	        \ii \,\,\text{erfc}\left(\ii\frac{-\ii \alpha + \gamma_{\aa}- \gamma_{\bb}-\ii \eta +\xi }{\sqrt{2}}\right)
	    \bigg]
	\Bigg]
	\label{haventlabeledit1}.
\end{align}
Taking the limit as $\eta\to0$ of $\mathcal{L}_{\aa\bb}^{\op \phi^2}$ yields
\begin{align}  
	\lim_{\eta\to0}\mathcal{L}&_{\aa\bb}^{\op \phi^2}  = 
    \frac{
        \lambda ^2 
        e^{ -\frac{ \beta  ^2}{2 \delta ^2}-\frac{1}{2} ( \gamma_{\aa} - \gamma_{\bb} )^2}
    }{32 \pi ^3 \delta  T^2 (  \beta   )}
	\int_0^\infty\frac{\d \xi}{ \xi ^2}
    \sinh \left(\frac{\xi  (  \beta   )}{\delta ^2}\right) 
	\notag\\&\times
    e^{-\ii \alpha  \xi 
        + \gamma_{\aa}  \xi
        - \gamma_{\bb}  \xi 
        -\frac{\xi ^2}{2 \delta ^2}
        -\frac{\xi ^2}{2}}
	\notag\\&\times
    \bigg[
        (\alpha \xi 
            +\ii\xi  ( \gamma_{\aa}  - \gamma_{\bb} -\xi) -\ii) 
        \notag\\&\quad\quad\times
	\,\text{erfc}\left(\frac{\alpha  +\ii  ( \gamma_{\aa} - \gamma_{\bb} -\xi )}{\sqrt{2}}\right)
	\notag\\&\quad
	-\sqrt{\frac{8}{ \pi} } \xi  e^{-\frac{1}{2} (\alpha  +\ii  ( \gamma_{\aa} - \gamma_{\bb} -\xi ))^2}
	\notag\\&\quad
        + e^{2 \xi  (\ii \alpha - \gamma_{\aa} + \gamma_{\bb} )}
        (\alpha  \xi +\ii \xi  ( \gamma_{\aa} - \gamma_{\bb} +\xi) +\ii)
        \notag\\&\quad\quad\times
        \,\text{erfc}\left(\frac{\alpha  +\ii  ( \gamma_{\aa} - \gamma_{\bb} +\xi )}{\sqrt{2}}\right)
    \bigg] \label{haventlabeledit2}.
\end{align}

Fig.~\ref{fig:allconvergencePlots} (e) illustrates the behaviour as $\eta$ decreases of the result of numerical integration over $\xi$.

\subsection{$\mathcal{M}$ non-local term in 3+1 dimensions}
In the following, we will derive  $\mathcal{M}^{\op\phi}$ and $\mathcal{M}^{\op\phi^2}$, then discuss the relevant differences between the two. We will see  that $\mathcal{M}^{\op\phi}$ has only regularizable divergences, while $\mathcal{M}^{\op\phi^2}$ exhibits persistent UV divergences.

\subsubsection{Linear coupling, $\mathcal{M}^{\op\phi}$}

For the usual linear detector, $\mathcal{M}$ is given by Eq.~\eqref{linearM}. We substitute into Eq.~\eqref{linearM} the  Wightman function in 3+1 dimensions \eqref{twopointLIN3+1}, the spatial profile \eqref{spatprof}, and the switching function \eqref{switchingFunct}. The integrals take a particualrly simple form under the same change of coordinates  \eqref{coords1} as in all previous calculations. In the case of $\mathcal{M}^{{\op\phi}}$, this change of coordinates also helps to de-nest the nested time integrals. This yields
\begin{align}    
    \mathcal{M}^{\op \phi} &=-
        e^{
            -\frac{ t_{\aa} ^2}{T^2}
            -\frac{ t_{\bb} ^2}{T^2}
            -\frac{ \bm x_{\aa} ^2}{\sigma ^2}
            -\frac{ \bm x_{\bb} ^2}{\sigma ^2}
        }
	\int_{-\infty}^{\infty}\!\!\!\d u
	\int_{-\infty}^{\infty}\!\!\!\d v
	\notag\\&\times 
	\int\!\d^3 \bm p
	\int\!\d^3 \bm q
    \Bigg[
    \frac{\lambda ^2 
        e^{
            -\frac{q  \bm x_{\aa} }{\sigma ^2}
            +\frac{q  \bm x_{\bb} }{\sigma ^2}
            -\frac{ t_{\aa}  v}{T^2}
            +\frac{ t_{\bb}  v}{T^2}
        }
    }{64 \pi ^5 \sigma ^6 \left(q^2-(v-\ii \epsilon )^2\right)}
	\notag\\&\quad
    +\frac{\lambda ^2 
        e^{-(
            -\frac{q  \bm x_{\aa} }{\sigma ^2}
            +\frac{q  \bm x_{\bb} }{\sigma ^2}
            -\frac{ t_{\aa}  v}{T^2}
            +\frac{ t_{\bb}  v}{T^2})
        }
    }{64 \pi ^5 \sigma ^6 \left(q^2-(v-\ii \epsilon )^2\right)}
    \Bigg]
	\notag\\&\times 
	    e^{
            -\frac{p^2}{2 \sigma ^2}
            -\frac{q^2}{2 \sigma ^2}
            -\frac{q  \bm x_{\aa} }{\sigma ^2}
            +\frac{q  \bm x_{\bb} }{\sigma ^2}
            +\frac{ t_{\aa}  u}{T^2}
            +\frac{ t_{\bb}  u}{T^2}
            -\frac{u^2}{2 T^2}
            -\frac{v^2}{2 T^2}
            +\ii u \Omega 
        }
\label{nonlocalMlin}.
\end{align}
The integrals in $u$, $\bm p$, and the angular parts of $\bm q$ can be evaluated in closed form. This results in 
\begin{align}    
    \mathcal{M}^{\op \phi} &=
    e^{
        -\frac{ t_{\aa} ^2}{2 T^2}
        +\frac{ t_{\aa}   t_{\bb} }{T^2}
        -\frac{ t_{\bb} ^2}{2 T^2}
        -\frac{T^2 \Omega ^2}{2}
        +\ii  t_{\aa}  \Omega
        +\ii  t_{\bb}  \Omega 
        -\frac{ \bm x_{\aa} ^2}{2 \sigma ^2}
        +\frac{ \bm x_{\aa}   \bm x_{\bb} }{\sigma ^2}
        -\frac{ \bm x_{\bb} ^2}{2 \sigma ^2}
    }
    \notag\\&\times
        \frac{\lambda ^2  T }{2 \pi ^2 \sigma  ( \bm x_{\aa} - \bm x_{\bb} ) }
	\int_{-\infty}^{\infty}\!\!\!\d v
	\int_0^\infty\!\!\!\d q
    \frac{ e^{
        -\frac{q^2}{2 \sigma ^2}
        -\frac{v^2}{2 T^2}
        }
    q }{ \left(-q^2+(v-\ii \epsilon )^2\right)}
    \notag\\&\times
    \sinh \left(\frac{q ( \bm x_{\aa} - \bm x_{\bb} )}{\sigma ^2}\right) 
    \cosh \left(\frac{v ( t_{\aa} - t_{\bb} )}{T^2}\right) .
\end{align}
\begin{figure*}[t]
    \centering
    \includegraphics[width=0.32\textwidth]{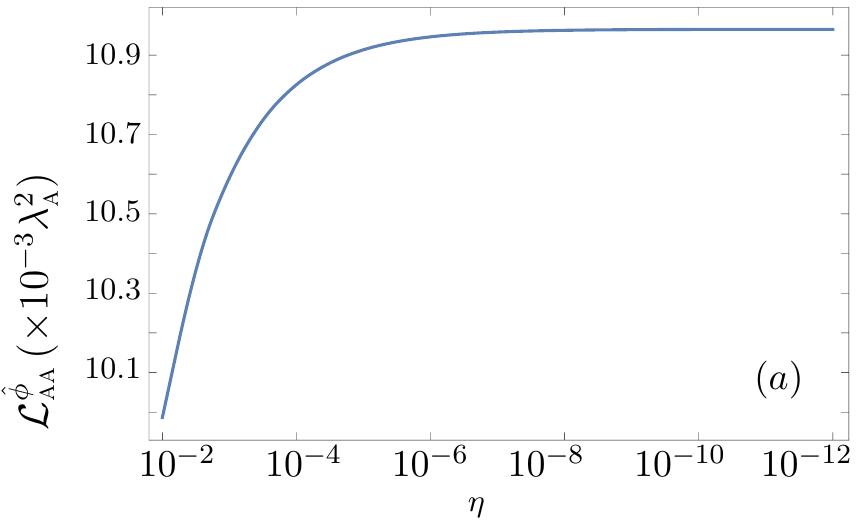}
    \includegraphics[width=0.32\textwidth]{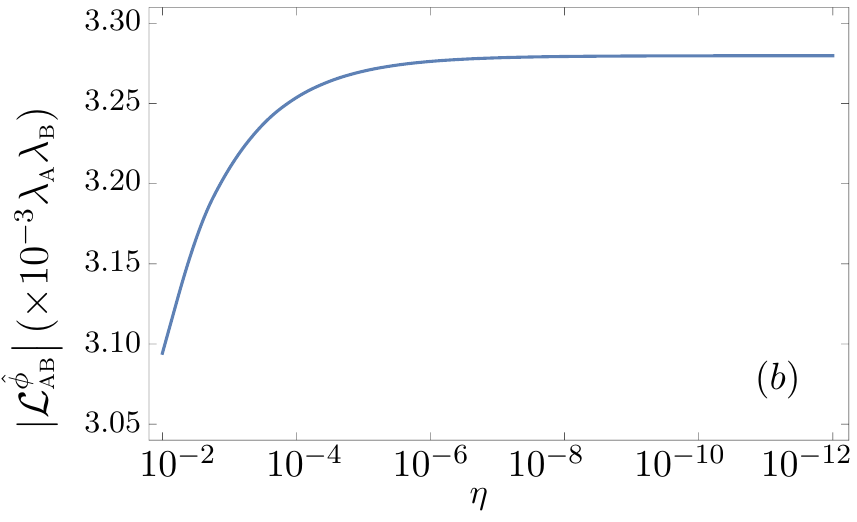}
    \includegraphics[width=0.32\textwidth]{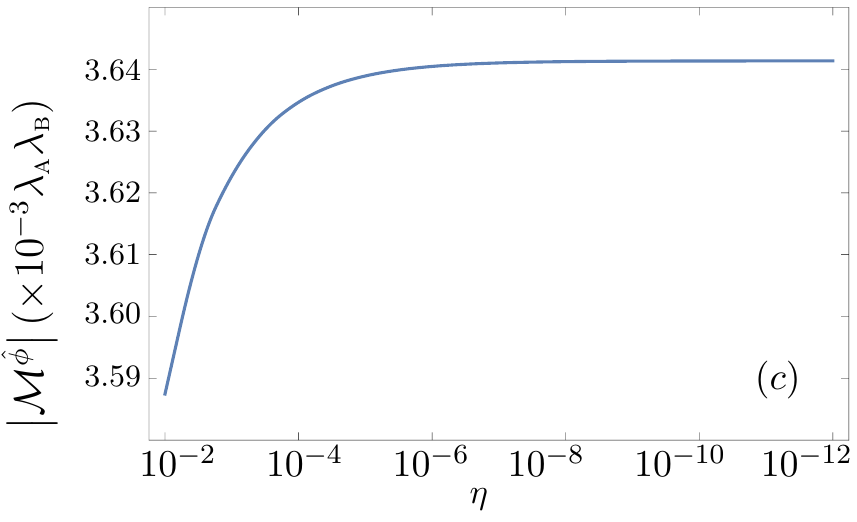}
    \\
    \includegraphics[width=0.32\textwidth]{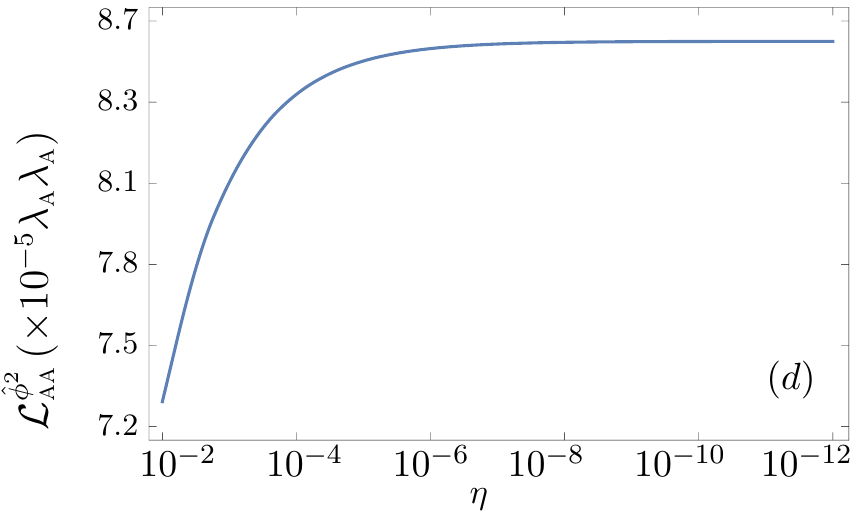}
    \includegraphics[width=0.32\textwidth]{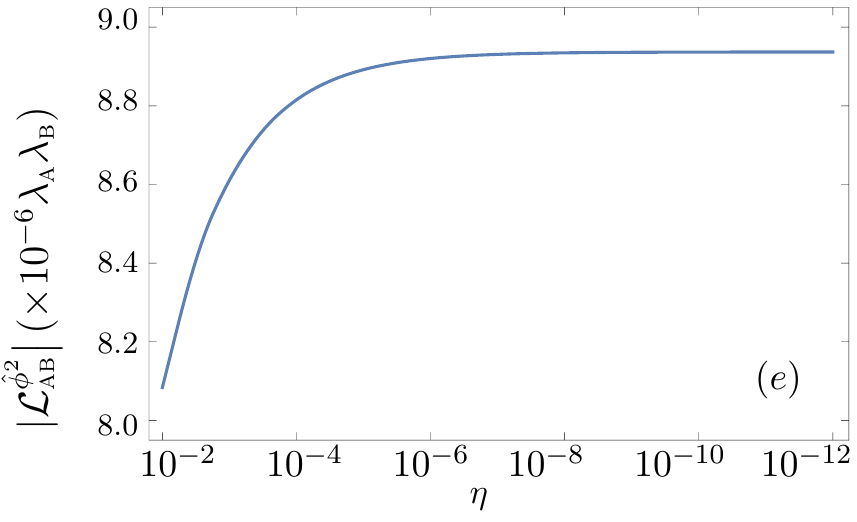}
    \includegraphics[width=0.32\textwidth]{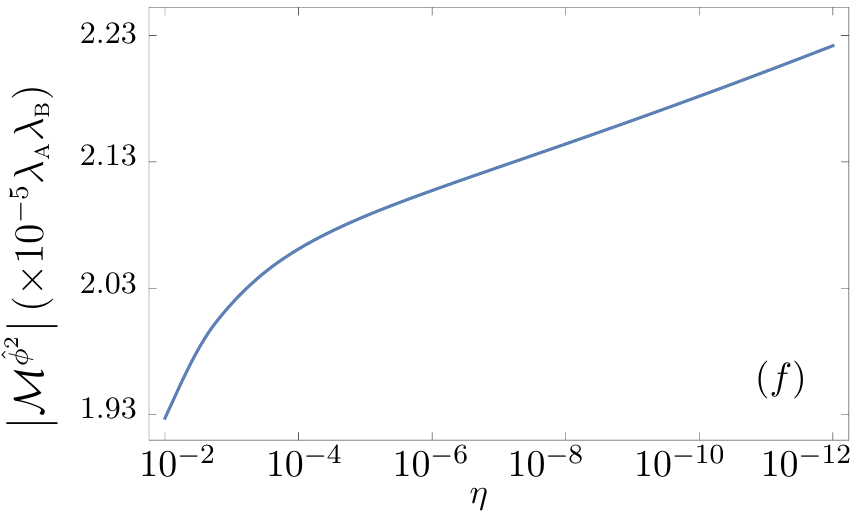}
    \caption{All plots illustrate the  behavior of relevant quantities as $\eta$ decreases on a log scale. Plots on the top row are for the usual (linearly coupled) UDW detector.  Plots on the bottom row are for the quadratically coupled UDW detector. All plots use  parameters $\alpha=1$, $\delta=1$, $\gamma_{\bb}-\gamma_{\aa}=4$, and $\beta=4$, where relevant. Note how all plots (a)-(e) indicate convergence, except  (f), which (in contrast to (c)) shows shows linear growth of $\mathcal{M}^{_{\op\phi^2}}$ on a logarithmic scale of $\eta$, and thus a logarithmic divergence   
    as $\eta\to 0$. }
    \label{fig:allconvergencePlots}
\end{figure*}
To obtain a closed form for the integral over $v$, we simplify $\mathcal{M}^{{\op\phi}}$ by choosing to switch on the detectors simultaneously within their co-moving frame, i.e. we make the simplifying additional assumption \mbox{$t_\aa -t_\bb=0$}. Under this assumption, $\mathcal{M}^{{\op\phi}}$ (in dimensionless parameters as shown in \ref{tab:parametersTable}) takes the form
\begin{align}    
\mathcal{M}^{\op \phi}&_{_{t_\aa =t_\bb}}=
    -\frac{\lambda ^2 }{16 \pi ^2 \delta    \beta }
	\int_0^\infty\!\!\!\d q
    \sinh \left(\frac{\xi   \beta  }{\delta ^2 }\right) 
    \notag\\&\times
    \Bigg[
        \bigg[
            2 \pi  \,\text{erfi}\left(\frac{\xi  T-\ii \eta  T}{\sqrt{2} T}\right)
            -2 \text{Ei}\left(\frac{(T \xi -\ii T \eta )^2}{2 T^2}\right)
    \notag\\&\quad\quad
            +\log \left(\frac{(\xi  T-\ii \eta  T)^2}{T^2}\right)
            -\log \left(\frac{T^2}{(\xi  T-\ii \eta  T)^2}\right)
    \notag\\&\quad\quad
            -4 \log (\xi  T-\ii \eta  T)
            +4 \log (T)
        \bigg]
        e^{-\frac{(\xi  T-\ii \eta  T)^2}{2 T^2}} 
    \notag\\&\quad
        +e^{-\frac{(\xi  T+\ii \eta  T)^2}{2 T^2}} 
        \bigg[
            2 \pi  \,\text{erfi}\left(\frac{\xi  T+\ii \eta  T}{\sqrt{2} T}\right)
    \notag\\&\quad\quad
            +2 \text{Ei}\left(\frac{(\ii T \eta +T \xi )^2}{2 T^2}\right)
            +\log \left(\frac{1}{(\xi  T+\ii \eta  T)^2}\right)
    \notag\\&\quad\quad
            +4 \log (-\xi  T-\ii \eta  T)
            -2 \log (\xi  T+\ii \eta  T)
        \bigg]
    \Bigg] 
    \notag\\&\times
    e^{-\frac{\alpha ^2 \delta ^2 T^2-4 \ii \alpha   \gamma_{\aa} \delta ^2 T^2+  \beta ^2 T^2 +\xi ^2 T^2}{2 \delta ^2 T^2}}
	\label{nonlocalMlinT0},
\end{align}
where is the principal value of the exponential integral function defined as
\begin{align}   
    \text{Ei}(z)&\coloneqq-\, \text{P.V.}\!\int_{-z}^{\infty } \frac{e^{-t}}{t} \, dt\label{expintegral}.
\end{align}

$\mathcal{M}^{\op\phi}$ is well behaved in the UV limit. If we remove the cutoff taking the limit \mbox{$\epsilon\rightarrow0$} (i.e., $\eta\to 0$), we obtain
\begin{align}   
\lim_{\eta\rightarrow0}\mathcal{M}^{\op \phi}_{_{t_\aa =t_\bb}}&=
    -\frac{\lambda ^2
    e^{
        -\frac{\alpha ^2}{2}
        +2 \ii \alpha   \gamma_{\aa} 
        -\frac{  \beta  ^2}{2 \delta ^2}}
    }{4 \pi  \delta    \beta   }
    \int_0^\infty\!\!\!\d \xi        
    e^{-\frac{\xi ^2}{2 \delta ^2}-\frac{\xi ^2}{2}}
    \notag\\&\times
    \ii\,\,\text{erfc}\left(\frac{\ii \xi }{\sqrt{2}}\right) 
    \sinh \left(\frac{\xi  (  \beta   )}{\delta ^2}\right) 
	\label{nonlocalMlinT0e0}.
\end{align}
The integral is convergent, and how the limit is reached as $\eta\rightarrow 0$ is shown in Fig.~\ref{fig:allconvergencePlots}.(c).

\subsubsection{Quadratic coupling, $\mathcal{M}^{\op\phi^2}$\label{quadcupM}}

For the quadratic detector, $\mathcal{M}^{\op \phi^2}$ is given by Eq.~\eqref{linearM}. We substitute into  Eq.~\eqref{linearM} the quadratic two-point correlator in 3+1 dimensions \eqref{twopointQUAD2}, the spatial profile \eqref{spatprof}, and the switching function \eqref{switchingFunct}. The traditional change of coordinates shown in Eq.~\eqref{coords1} simplifies $\mathcal{M}^{\op \phi^2}$. The result of these substitutions is
\begin{align}    
    \mathcal{M}^{\op \phi^2} &=-
        e^{
            -\frac{ t_{\aa} ^2}{T^2}
            -\frac{ t_{\bb} ^2}{T^2}
            -\frac{ \bm x_{\aa} ^2}{\sigma ^2}
            -\frac{ \bm x_{\bb} ^2}{\sigma ^2}
        }
	\int_{-\infty}^{\infty}\!\!\!\d u
	\int_{-\infty}^{\infty}\!\!\!\d v
	\notag\\&\times 
	\int\!\d^3 \bm p
	\int\!\d^3 \bm q
    \Bigg[
    \frac{\lambda ^2 
        e^{
            -\frac{q  \bm x_{\aa} }{\sigma ^2}
            +\frac{q  \bm x_{\bb} }{\sigma ^2}
            -\frac{ t_{\aa}  v}{T^2}
            +\frac{ t_{\bb}  v}{T^2}
        }
    }{128 \pi ^5 \sigma ^6 \left(q^2-(v-\ii \epsilon )^2\right)^2}
	\notag\\&\quad
    +\frac{\lambda ^2 
        e^{-(
            -\frac{q  \bm x_{\aa} }{\sigma ^2}
            +\frac{q  \bm x_{\bb} }{\sigma ^2}
            -\frac{ t_{\aa}  v}{T^2}
            +\frac{ t_{\bb}  v}{T^2})
        }
    }{128 \pi ^5 \sigma ^6 \left(q^2-(v-\ii \epsilon )^2\right)^2}
    \Bigg]
	\notag\\&\times 
	    e^{
            -\frac{p^2}{2 \sigma ^2}
            -\frac{q^2}{2 \sigma ^2}
            -\frac{q  \bm x_{\aa} }{\sigma ^2}
            +\frac{q  \bm x_{\bb} }{\sigma ^2}
            +\frac{ t_{\aa}  u}{T^2}
            +\frac{ t_{\bb}  u}{T^2}
            -\frac{u^2}{2 T^2}
            -\frac{v^2}{2 T^2}
            +\ii u \Omega 
        }
\label{nonlocalMquad}.
\end{align}
The integrals over $u$, $\bm p$, and the angular part of $\bm q$ can be evaluated in closed form. We can furthermore  employ the simplifying assumption that the two detectors are switched on simultaneously, which results in 
\begin{align}    
& \mathcal{M}_{_{t_\aa =t_\bb}}^{\op\phi^2}=
	-\frac{\lambda ^2 T 
	    e^{-\frac{T^2 \Omega ^2}{2}+2\ii  t_{\aa}  \Omega -\frac{ (\bm x_{\aa} -\bm x_{\bb} )^2  }{2\sigma ^2}}
	}{4 \pi ^4 \sigma  ( \bm x_{\aa} - \bm x_{\bb} ) }
	\notag\\&\times
    \int_0^\infty\!\!\!\d q
	\int_{-\infty}^{\infty}\!\!\!\d v
    \frac{q\sinh \left(\frac{q ( \bm x_{\aa} - \bm x_{\bb} )}{\sigma ^2}\right) }{\left(q^2-(v-\ii \epsilon )^2\right)^2} 
    e^{-\frac{q^2}{2 \sigma ^2}-\frac{v^2}{2 T^2}}
	\label{quadMpos}.
\end{align}
To obtain a closed from for the integral over $v$, we operate as in the linear case and simplify by choosing to switch on the detectors simultaneously within their co-moving frame, setting \mbox{$t_\aa -t_\bb=0$}. The resulting semi-closed form we write as 
\begin{align}    
    \mathcal{M}_{_{t_\aa =t_\bb}}^{\op\phi^2}&=
		-\frac{\lambda ^2 
            e^{
		    -\frac{\alpha ^2}{2}
		    +2 \ii \alpha  \gamma_{\aa}
		    -\frac{ \beta ^2}{2 \delta ^2}
            }
		}{64 \pi ^4 \delta   T^3  \beta  }
		\int_0^\infty\!\!\d \xi \,
		\text{G}
		\left(\xi\right)\label{nonlocalquad}
\end{align}
after carrying out the integral over $v$. The details of $\text{G}\left(\xi\right)$ can be found in appendix \ref{ap:QuadM}, concretely Eq.~\eqref{finalIntegrandQuad}.

\subsection{Divergences in the quadratic model}

Unlike the linear model, the non-local term $\mathcal{M}^{\op\phi^2}$ is not free of UV divergences, despite the fact that the detector has a smooth switching and a Gaussian spatial smearing, and despite the renormalization process that removed the single-detector divergences. Concretely, the integral in  \eqref{nonlocalquad} is logarithmically divergent with the UV cutoff scale, as  illustrated  in Fig.~\ref{fig:allconvergencePlots}.(f).

To gain  insight on the logarithmic divergence in $\mathcal{M}^{\op\phi^2}$  we examine the integrand $G(\xi)$, defined in the appendix equation \eqref{finalIntegrandQuad}. 

We begin by noticing that $G(\xi)$ has the limit $\eta\to0$
\begin{align}
    \lim_{\eta\to0}G&(\xi)\!=
    \frac{4}{\xi ^2}
    e^{-\frac{\left(\delta ^2+1\right) \xi ^2}{2 \delta ^2}} 
    \sinh \left(\frac{\xi   \beta }{\delta ^2}\right)
    \\&\times\notag
    \Bigg[-\sqrt{2 \pi } e^{\frac{\xi ^2}{2}} \xi +\ii \pi  \left(\xi ^2+1\right)  \,\text{erfc}\left(\frac{\ii\xi }{\sqrt{2}}\right) \Bigg] 
    .
\end{align}
Expanding in Laurent series and keeping the leading order $\mathcal{O}(\xi^{-1})$ results in the UV divergent term
\begin{align}    
    \lim_{\eta\to0}G&_{\mathcal{M}_{_{t_\aa =t_\bb}}^{\op\phi^2}}\sim\frac{4 \ii \pi   \beta  }{\delta ^2 \xi }.
\end{align}

This divergence is peculiar due to the fact that it shows up only in the two-detector model, in spite of the fact that the vacuum excitation probability for a quadratic detector is finite   \cite{hummer} as discussed in section \ref{quadVEP}. Thus, while a single  quadratically coupled detector does not require additional UV regularization, a cutoff is required for certain quantities describing detector pairs, of which the $\cal M$ term in \eqref{nonlocalquad} is an example.

We would like to emphasize that these are \textit{persistent} UV divergences, that is, they are present regardless of the use of smooth switching functions and spatial profiles (for example, in this case we have used Gaussian functions for both). Moreover, these divergences appear after renormalization of the zero-point energy and at the same order in perturbation theory at which the single detector dynamics is regular.

\section{Harvesting Correlations 
\label{harvesting}}

The analysis of the correlation terms $\mathcal{L}_\textsc{ab}$ and $\mathcal{M}$ in both the linear and quadratic models is necessary to explore the entanglement structure of the field through particle detectors.

In this section we will study two types of correlations that the two detector models can harvest from the field vacuum: a) those measured by the mutual information, which quantifies both classical and quantum correlations \cite{NandChuang}, and b) the  negativity, which is a faithful entanglement measure for bipartite two-level systems \cite{Vidal:2002aa}. These two types of correlation harvesting were studied for the linear model in \cite{Pozas:2015aa}. Here we will compare these results with the predictions for the quadratic model.

\subsection{Entanglement Harvesting
\label{entharvesting}}
 
\begin{figure*}[ht]
    \centering
    \includegraphics[width=\textwidth]{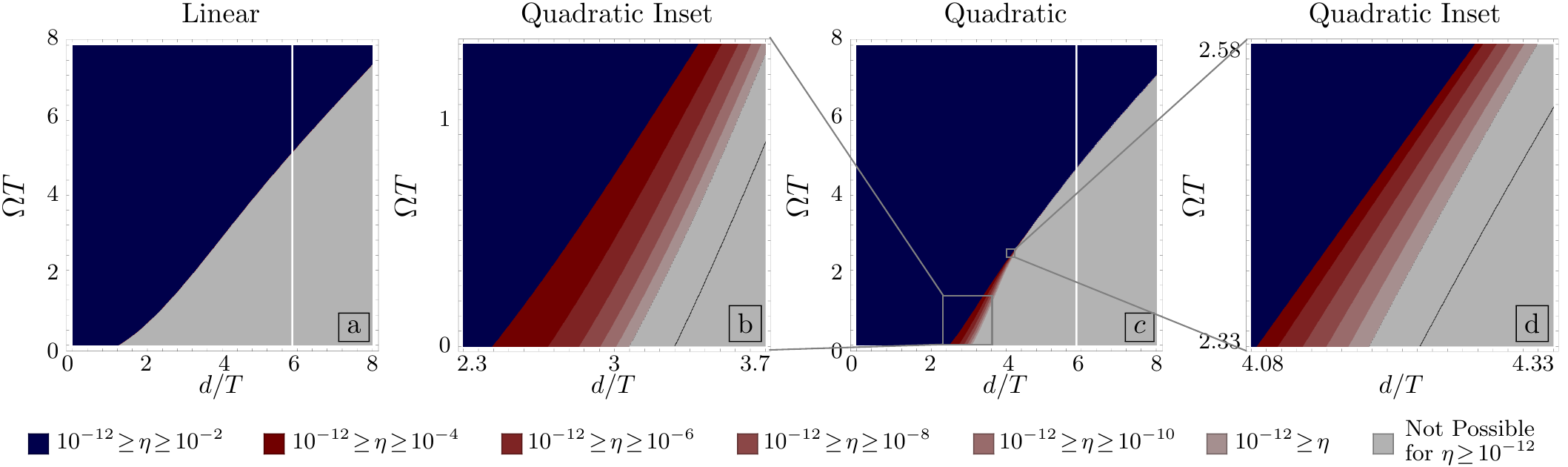}
    \caption{Entanglement harvesting is possible (darker regions colored red or blue) for any detector separation $d$ given a sufficiently large detector gap $\Omega$. These plots show various detector cutoffs, $\epsilon=\eta/T$, ranging from $\eta=10^{-2}$ to $\eta=10^{-12}$. Note the cutoff does not make a significant difference for the linear model (no visible red regions), while there is a marked increase in the harvesting region for smaller cutoffs (see inset (b) and (c)). Both plots use  parameters $\delta=1$ and $\gamma_{\bb}-\gamma_{\aa}=0$. The vertical white line shows the light cone. The dark lines in the insets are an extrapolation indicating the  location of where harvesting is no longer possible for a cutoff $\eta=10^{-29}$ (corresponding to setting the cutoff scale to the Planck frequency, as described at the end of section \ref{entharvesting}).}
    \label{fig:Entanglementharv}
\end{figure*}

We consider first the harvesting of entanglement from the vacuum and we quantify it with the negativity acquired between the two (initially uncorrelated) detectors through their interactions with the field while remaining spacelike separated.

Negativity of a bipartite state $\rho$ is an entanglement monotone defined as the sum of the negative eigenvalues of the partial transpose of $\rho$ \cite{Vidal:2002aa}:
\begin{align}
    \mathcal{N}(\rho) =\!\!\! \sum_{\lambda_{i}\in\, \sigma\left[\rho^{\Gamma_{\!A}}\right]}\!\! \frac{|\lambda_{i}|-\lambda_{i}}{2} ,
\end{align}
where  $\rho^{\Gamma_A}$ denotes the partial transpose of $\rho$ with respect to subsystem A.

As seen, for instance, in \cite{Pozas:2015aa}, the negativity can be expressed in terms of the vacuum excitation probability, $\mathcal{L}_{\mu\mu}$,  of each detector and the non-local term $\mathcal{M}$ in \eqref{nonlocalquad}. Concretely, it is given by

\begin{equation}
\mathcal{N}=\max\left[\mathcal{N}^{(2)}, 0\right] +\mathcal{O}(\lambda^2),
\end{equation}
where
\begin{align}    
    \mathcal{N}^{(2)}\!=
    -\frac{1}{2}
    \left( \!
        \mathcal{L}_{\aa\aa} \!
        +\mathcal{L}_{\bb\bb} 
        -\sqrt{
            \left(\mathcal{L}_{\aa\aa} -\mathcal{L}_{\bb\bb} \right)^2+4\left|\mathcal{M}\right|^2
        }
    \right).
    \label{N-2}
\end{align}
When both detectors are identical (i.e. they have the same spatial profile, switching function, coupling strength, and detector gap), Eq.~\eqref{N-2} becomes
\begin{equation}
\mathcal{N}^{(2)}=|\mathcal{M}|-\mathcal{L}_{\mu\mu} \label{negat2equal}
\end{equation}
from which we can justify the usual argument that entanglement emerges as a competition between the non-local contribution  $\mathcal{M}$ and the noise associated  to the vacuum excitation probability for each detector  \cite{Reznik:2005aa,Pozas:2015aa}. 

Figure \ref{fig:Entanglementmag} shows the behavior of the negativity with the spatial separation of the detectors, for the linear and quadratic case and a range of detector cutoffs.

Recall that the term $\mathcal{M}$ is UV divergent in the quadratic model, therefore to compute a physically meaningful value for the negativity further regularization and eventual renormalization would be required. However for a fixed UV-cutoff scale, it is possible to get an estimate of the quadratic model performance to harvest entanglement relative to the linear model by computing entanglement harvesting for both models applying the same UV-cutoff scale. What is more, studying how negativity changes as we start increasing the cutoff scale will help us see how the UV divergence of $\mathcal{M}$ impacts entanglement harvesting.

As seen in Fig.~\ref{fig:Entanglementmag}, the magnitude of entanglement harvesting increases linearly with the logarithm of the cutoff, which is not surprising since the two-detector quadratic  model suffers a logarithmic UV divergence. This implies that  there would always exist a value for the cutoff scale so that harvesting is possible at any distance, regardless of the detector gap. It would also imply that for large enough cutoff frequencies we could always `harvest' more entanglement with the quadratic model than for the linear model.

One can therefore ask the following question:  is there any finite value of the cutoff scale that we could take in order to give some physical meaning to the finite cutoff results?

Unlike the linear model---which has been shown to capture the fundamental features of the atom-light interaction \cite{Salton:2015aa,Edu:2013aa}--- the quadratic model does not have a direct comparison with something as simple as the atom-light interaction mechanism (maybe one could think of non-linear optical media \cite{scully_zubairy_1997}, but that is perhaps a stretch). Recall, however, that we do not use the quadratic Unruh-DeWitt model to necessarily reproduce the physics of a particular experimentally motivated setup. Our motivation to explore this model is double: a) probe the field with a different model to show model independence/dependence of harvesting phenomena and b) advance towards the fermionic model (which is a quadratic model that does indeed have physical motivation) where the study of field entanglement remains still full of open questions.

The fact that this model cannot be connected with something as simple as an atom interacting with light, makes it difficult to motivate a choice of cutoff. However, if we were to take the result for a finite value of the cutoff scale seriously, and thus if we were to choose some physically motivated cutoff, we could compare the two models when such a cutoff is taken to be the Planck Frequency. In this scenario, the dimensionless cutoff parameter $\eta$ can be written as \mbox{$\eta=\frac{1}{k_{\textsc{p}} T}$}, where $\mathrm{k}_{\textsc{p}}$ is the Planck frequency. If we consider scales for the detector gap $\Omega$ to be commensurate with the energy of the first transition of Hydrogen \mbox{$\Omega_\textsc{h}\approx 10^{15}$ s$^{-1}$}, then $k_\textsc{p}=10^{29}\Omega$. If we set $T\approx \Omega_\textsc{h}^{-1}$ (which means that $\alpha\approx 1$ represents the case of a Hydrogen atom) the cutoff associated with the Planck time is then $\eta=10^{-29}$. We can extrapolate the results in Fig.~\ref{fig:Entanglementharv} to the Planck scale. We show these results also in Fig~\ref{fig:Entanglementharv}, as the thin black lines. These plots illustrate the slow logarithmic nature of the divergences, which  makes the study of negativity still meaningful for low energies with a quadratic detector and does not get significantly qualitatively modified even if the cutoff is transplanckian.

\subsection{Harvesting Mutual Information}

\begin{figure*}[ht]
    \centering
    \includegraphics[width=0.33\textwidth]{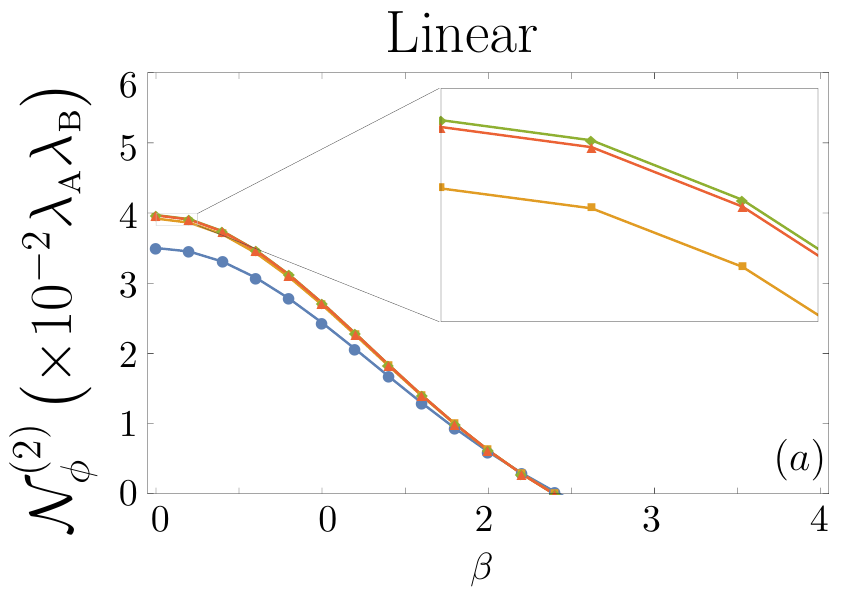}
    \includegraphics[width=0.33\textwidth]{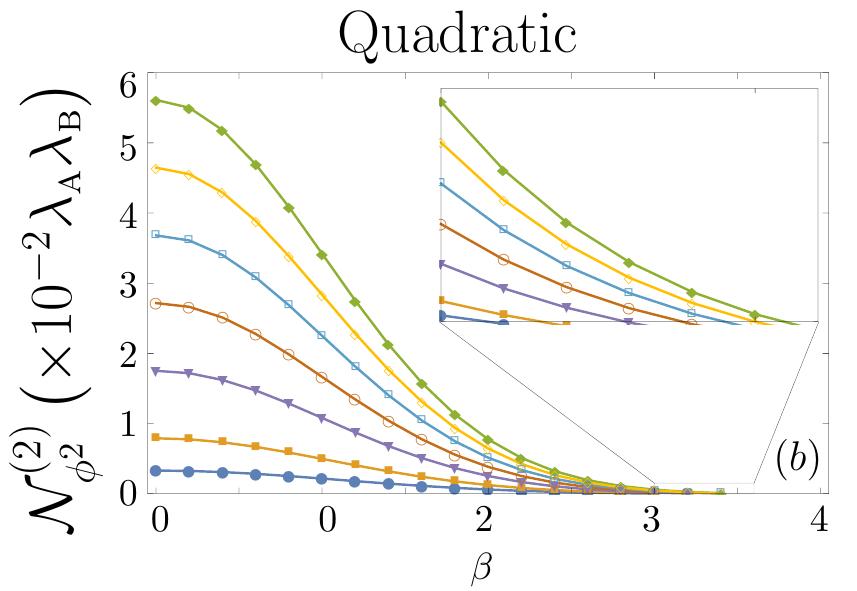}\\
    \includegraphics[width=0.66\textwidth]{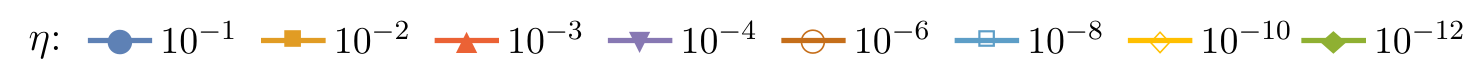}
    \caption{The magnitude of entanglement harvested is dependent on the cutoff chosen. For  the same cutoff, either the linear or quadratic models may harvest more entanglement, although the linear model quickly converges to a fixed value, while the quadratic model grows logarithmically with decreasing cutoff. Thus, a cutoff can always be chosen sufficiently high such that the quadratic model harvests more entanglement for a given set of parameters. Here, the top row shows plots of leading order negativity with increasing spatial distance for the linear model, while the bottom row shows the same thing for the quadratic model. The insets in each plot who where the negativity goes to zero. All plots use  parameters $\delta=1$ and $\gamma_{\bb}-\gamma_{\aa}=0$, while (a) and (c) show harvesting for detector gap $\alpha=0$ and (b) and (d) show  $\alpha=1$. }
    \label{fig:Entanglementmag}
\end{figure*}

A way around the problems associated to the UV-divergent nature of $\mathcal{M}$ is to look at UV-safe quantities. Namely, it is possible to find quantifiers of correlations that are, by construction, UV-safe for the quadratic model.  One such figure of merit is the mutual information. 

The mutual information $I(\rho_{\aa\bb})$ between   two detectors  quantifies the amount of uncertainty   about one detector that is eliminated if some information about the state of the other is revealed \cite{NandChuang}. Thus, it constitutes a faithful measure of correlations (regardless if they are classical or quantum).

\begin{figure}[h]
    \centering
    \includegraphics[width=\columnwidth]{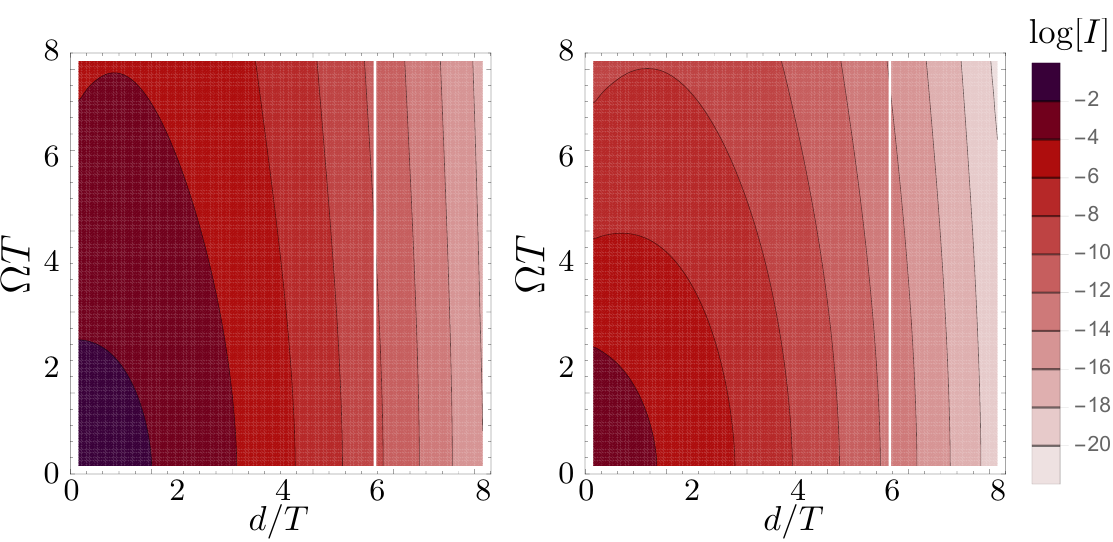}
    \caption{Linear mutual information harvesting (left) is greater in magnitude that quadratic mutual entanglement harvesting (right). Legend indicates Log base ten of the mutual information, $I$. 
    Both plots use  parameters  $\delta=1$ and $\gamma_{\bb}-\gamma_{\aa}=0$. The thick black line shows the light cone. }
    \label{fig:MutualInfo}
\end{figure}

In general, for composite quantum system consisting of two subsystems A and B, the mutual information given by
\begin{equation}
I(\rho_{\aa\bb})=S(\rho_\aa)+S(\rho_\bb)-S(\rho_{\aa\bb}),
\end{equation}
where $\rho_\nu=\text{Tr}_{\mu}(\rho_{\nu\mu})$ is the partial trace of $\rho_{\nu\mu}$ with respect to subsystem $\mu\in\{A,B\}$ and $S$ is the von Neumann entropy given by $S(\rho)=-\text{Tr}(\rho\log\rho)$.

For a density  matrix of the form \eqref{densitymatrix}, the mutual information is given by \cite{Pozas:2015aa} 
\begin{align}    
\notag I(\rho_{\aa\bb} )=&\,\mathcal{L}_{+} \log (\mathcal{L}_{+})+\mathcal{L}_{-} \log (\mathcal{L}_{-})\\
&-\mathcal{L}_{\aa\aa}  \log (\mathcal{L}_{\aa\aa} )-\mathcal{L}_{\bb\bb}  \log (\mathcal{L}_{\bb\bb} )+\mathcal{O}(\lambda_\nu^4) 
\label{MI-2}
\end{align}
where  
\begin{equation}
\mathcal{L}_\pm=\frac{1}{2}\left(\mathcal{L}_{\aa\aa} +\mathcal{L}_{\bb\bb} \pm\sqrt{(\mathcal{L}_{\aa\aa} -\mathcal{L}_{\bb\bb} )^2+4 \left|\mathcal{L}_{\aa\bb} \right|^2}\right).
\end{equation}
Note how $I(\rho_{\aa\bb} )$ is not dependent on the divergent $\mathcal{M}$ term at leading order in perturbation theory. Hence, the mutual information is finite without any further regularization. Because of this, it provides a UV-cutoff independent sense of the harvesting of correlations from the vacuum, and can also be compared with previous results for linear detectors in \cite{Pozas:2015aa}, making it a relevant figure of merit for the comparison in this article. 

Fig.~\ref{fig:MutualInfo} shows the behavior of the Mutual information with spatial and temporal separation of the detectors for both the linear and quadratic case where the other parameters are the same as those used in Fig \ref{fig:Entanglementharv}. First, we observe something that was already present in previous liteatura on linear detectors \cite{Pozas:2015aa}: Unlike entanglement, the mutual information harvesting can be performed at any distance and detector gap, albeit less efficiently as the distance (or the detector gap) increases, a feature that comes from the fact that the detectors are harvesting classical correlations as well as quantum correlations.

From Fig.~\ref{fig:MutualInfo} we observe that the linear detector can harvest more entanglement and for further distances than the quadratic detectors. This can in turn be used to assess the scale at which the soft cutoff model introduced in the study of negativity fails to capture the behaviour of UV-safe measures of correlations: as illustrated in Fig \ref{fig:Entanglementmag}, for cutoff scales that are of the order of $\eta \gtrsim 10^{-6}$, the linear model can harvest more entanglement than the quadratic model in the parameter region studied. This might suggest that comparison of negativity between the two models can be trusted only for cutoffs above \mbox{$\eta =10^{-6}$}.

\section{Conclusion
\label{conclusion}}

We have studied further the behaviour of particle detectors quadratically coupled to scalar fields introduced in \cite{Takagi:1985aa,Takagi:1986aa}  and renormalized in \cite{hummer}. In particular we have focused on the case of a pair of particle detectors harvesting entanglement from a scalar field, a case previously studied only for linear detectors   \cite{Reznik:2005aa,Reznik:2003aa,Pozas:2015aa}. Understanding the harvesting of correlations from quadratic couplings is a necessary step in order to compare the entanglement that can be harvested from fermionic and bosonic fields, since the former only couple to particle detectors quadratically \cite{hummer,Takagi:1985aa,Takagi:1986aa,Jorma:2016aa}. Our motivation to explore this model is twofold: a) probe the field with a different particle detector model to show model independence/dependence of harvesting phenomena and b) provide a model that can be compared on equal footing for bosonic and fermionic fields (for which the coupling necessarily has to be quadratic).

Perhaps the most remarkable finding of our investigation of   harvesting with a quadratic detector is the appearance of a new logarithmic UV divergence at leading order in the two-detector setup. Notably, this divergence  remains even when the Hamiltonian is normal-ordered, and even when the switching functions and spatial profile are smooth functions. This is in stark contrast with the linear case where smooth smearing \cite{Jorma:2006aa}  or switching \cite{Jorma:2008aa,satz:2007aa} were enough to guarantee the UV regularity of the model.

We emphasize that a single detector, at the same order in perturbation theory, does not present this kind of divergence.  Curiously,  the UV divergence is only present in a particular kind of term, namely that  responsible for the entanglement of the two detectors. This divergence is easily parametrized via  a UV cutoff.

Once this was established, analysis  and comparison with the linear model can be carried out.  We proceeded in two different ways. First, using negativity to study entanglement harvesting. We discussed whether a finite value  of the UV cutoff scale   allows for fair comparison of the entanglement harvesting ability of the quadratic and the linear couplings. Following this, we  found measures of harvested correlations that are UV-safe. In particular we   showed that the harvested mutual information from the field vacuum is UV safe. It therefore constitutes a better figure of merit to compare the harvesting of correlations from the vacuum without need for further regularization.

Understanding the particulars of entanglement harvesting with  bosonic quadratic coupling is important in order to properly answer questions about fermionic fields where the study of field entanglement remains   full of open questions \cite{PhysRevA.74.032326,PhysRevA.80.042318,PhysRevA.81.032320,PhysRevA.81.052305,PhysRevD.82.045030,PhysRevA.82.042332,PhysRevD.82.064006,PhysRevA.83.052306,Montero:2011aa,PhysRevA.84.012337,Montero:2011ab,PhysRevA.84.042320,PhysRevA.84.062111,PhysRevA.85.016301,PhysRevA.85.016302,PhysRevD.85.025012,PhysRevA.85.024301,PhysRevA.87.022338}. A comparison of bosonic and fermionic entanglement harvesting on equal footing requires knowledge of the model-dependence  entanglement harvesting, specifically the difference between linear and quadratic coupling, as the latter is necessarily present in the fermionic case. The entanglement structure of the fermionic vacuum remains an interesting open question, one we are now prepared to address using the results we have obtained.

 \section*{Acknowledgements}

The authors would like to thank Jose De Ram\'on for helpful  discussions. E.M-M. and R.B.M. acknowledge the support of the Natural Sciences and Engineering Research Council of Canada NSERC programme. E.M-M also acknowledges the support of the Ontario Early Research Award.

\onecolumngrid
\appendix\section{Calculation of the quadratic two-point function 
\label{ap:wick}}

Here we demonstrate that
\begin{align}  
	W^{\op\phi^2} (t,\bm{x},t'\!,\bm{x}') 	\!	= 2\left(W^{\op\phi} (t,\bm{x},t'\!,\bm{x}')\right)^2 \label{AWickTrick},
\end{align}
as asserted in section \ref{LinCupSect} and \ref{QuadCupSect},  
where $W^{\op\phi^2}$ and $W^{\op\phi}$ are defined in Eqs. \eqref{twopointQUAD} and \eqref{twopointLIN}, respectively.  

The relationship between an operator $\op A$ and its normal ordered version is given by:
\begin{align}  
	 : \! \op A \! : \,= \op A - \bra{0} \op A \ket{0} \label{normaldef}.
\end{align}
Using this identity, $W^{\op \phi^2}$ can be rewritten as
\begin{align}  
    W^{\op\phi^2}&(t,\bm{x},t'\!,\bm{x}') = 
        \bra{0}\!: \!\op \phi^2(t,\bm{x}) \!:\,:\!\op \phi^2(t'\!,\bm{x}') \!:\!\ket{0}= 
        \bra{0} \op \phi^2(t,\bm{x}) \op \phi^2(t'\!,\bm{x}') \ket{0}
        - \bra{0} \op \phi^2(t,\bm{x}) \ket{0} \bra{0} \op \phi^2(t'\!,\bm{x}') \ket{0}.\label{thingBeforePhiCommies}
\end{align}
The first term of $W^{\op\phi^2}(t,\bm{x},t'\!,\bm{x}')$ can be simplified. To do so, we will write the field operator as $\op \phi=\op \phi^++\op \phi^-$, where $\op \phi^+$ and $\op \phi^-$ are defined as
\begin{align}  
	\op\phi^+ (\bm{x},t) =
		\int  \frac{\d^n \bm{k} \, 
		e^{-\epsilon|\bm{k}|/2}}{\sqrt{2(2\pi)^n|\bm{k}|}}\,
		    \hat{a}^\dagger_k \,
		    e^{\ii (|\bm{k}|t-\bm{k}\cdot\bm{x})}     \qquad
    \op\phi^- (\bm{x},t) =
		\int  \frac{\d^n \bm{k} \, 
		e^{-\epsilon|\bm{k}|/2}}{\sqrt{2(2\pi)^n|\bm{k}|}} \,
		    \hat{a}^{\phantom{\dagger}}_k \,
		    e^{-\ii (|\bm{k}|t-\bm{k}\cdot\bm{x})}
\end{align}
which satisfy the commutation relation
\begin{align}  
    \big[ \op\phi^- (\bm{x}_\mu,t_\mu) ,  \op\phi^+ (\bm{x}_\nu,t_\nu) \big]=\mathcal{C}_{\mu\nu}\openone\label{comrelation1}
\end{align}
where $\mathcal{C}_{\mu\nu}\in \mathbb{C}$ is given by
\begin{align}  
    \mathcal{C}_{\mu\nu}=&
		\int  \frac{\d^n \bm{k} \, e^{-\epsilon|\bm{k}|/2}}{2(2\pi)^n|\bm{k}|}
		e^{\ii (|\bm{k}|(t_\nu-t_\mu)-\bm{k}\cdot(\bm{x}_\nu-\bm{x}_\mu))}
\end{align}

Using the notation $\op\phi_\mu\equiv\op\phi(\bm{x}_\mu,t_\mu)$, a scalar field vacuum four point function  $\bra{0} \op \phi_1  \op \phi_2 \op \phi_3 \op \phi_4 \ket{0}$ can be rewritten as
\begin{align}  
        \bra{0}& \op \phi_1       \op \phi_2      \op \phi_3      \op \phi_4      \ket{0}=
        \bra{0}  \op \phi^-_1  \! \op \phi^-_2 \! \op \phi^+_3 \! \op \phi^+_4 \! \ket{0}+
        \bra{0}  \op \phi^-_1  \! \op \phi^+_2 \! \op \phi^-_3 \! \op \phi^+_4 \! \ket{0}\label{peeeeepobastard},
\end{align}
where,   to remove all vanishing summands, we have used that
\begin{equation}
\op\phi_\mu^-\ket{0}=\bra{0}\op\phi_\nu^+=0,
\end{equation}
together with the fact that only summands with as many $\op\phi^+$ as $\op\phi^-$ give a non-vanishing vacuum expectation.

Using \eqref{comrelation1}, we can write the first summand in Eq.~\eqref{peeeeepobastard} as
\begin{align}  
        \bra{0} \op \phi^-_1  \op \phi^-_2 \op \phi^+_3 \op \phi^+_4 \ket{0}=
        \mathcal{C}_{23}\mathcal{C}_{14} + \mathcal{C}_{13}\mathcal{C}_{24}
\end{align}
and the second as 
\begin{align}  
    \bra{0} \op \phi^-_1  \op \phi^+_2 \op \phi^-_3 \op \phi^+_4 \ket{0}
    =\mathcal{C}_{12}\mathcal{C}_{34} .
\end{align}
Thus \eqref{peeeeepobastard} can be written as
\begin{align}  
        \bra{0}& \op \phi_1  \op \phi_2 \op \phi_3 \op \phi_4 \ket{0}=
        \mathcal{C}_{23}\mathcal{C}_{14} + \mathcal{C}_{13}\mathcal{C}_{24} + \mathcal{C}_{12}\mathcal{C}_{34}.\label{OMGtheDeltas}
\end{align}
From \eqref{comrelation1}, we see that we can rewrite the $\mathcal{C}_{\mu\nu}$ coefficients as 
\begin{align}  
    \mathcal{C}_{\mu\nu}=\bra{0}\big[ \op\phi_\mu^-,\op\phi_\nu^+ \big]\ket{0}=\bra{0} \op\phi_\mu^-\op\phi_\nu^+\ket{0}=\bra{0} \op\phi_\mu\op\phi_\nu\ket{0}.
\end{align}
This allows us to rewrite Eq.~\eqref{OMGtheDeltas} as
\begin{align}  
        \bra{0}& \op \phi_1  \op \phi_2 \op \phi_3 \op \phi_4 \ket{0}=
        \bra{0} \op \phi_1  \op \phi_2\ket{0}\bra{0} \op \phi_3 \op \phi_4 \ket{0}
        +\bra{0} \op \phi_2  \op \phi_3\ket{0}\bra{0} \op \phi_1 \op \phi_4 \ket{0} 
        +\bra{0} \op \phi_1  \op \phi_3\ket{0}\bra{0} \op \phi_2 \op \phi_4 \ket{0}.
\end{align}

To apply this identity to \eqref{thingBeforePhiCommies}, we set $\op\phi_1=\op\phi_2=\op\phi(t,\bm{x})$ and $\op\phi_3=\op\phi_4=\op\phi(t',\bm{x}')$. Then, the first summand in \eqref{thingBeforePhiCommies} becomes
\begin{align}  
        \bra{0}& \op \phi^2(t,\bm{x})               \op \phi^2(t'\!,\bm{x}') \ket{0}=
        \bra{0}  \op \phi^2(t,\bm{x})\ket{0}\bra{0} \op \phi^2(t'\!,\bm{x}') \ket{0}
        +2\left(\bra{0} \op \phi(t,\bm{x}) \op \phi(t'\!,\bm{x}')\ket{0}\right)^2 .
\end{align}
Which allows \eqref{thingBeforePhiCommies} to be written as
\begin{align}  
    W^{\op\phi^2}(t,\bm{x},t'\!,\bm{x}') = 
        2\left(\bra{0} \op \phi(t,\bm{x}) \op \phi(t'\!,\bm{x}')\ket{0}\right)^2,
         \label{thing1}
\end{align}
which is \eqref{AWickTrick}.

\section{Convolution 
\label{ap:convolution}}

In this appendix, we will find a closed form for the following integral
\begin{align}
    f_m& \coloneqq \int_{-\infty}^{\infty}\!\!\!\!\!\d v\,
	\frac{
	    e^{v\frac{(t_{\aa}-t_{\bb})}{T^2}
	        -\frac{v^2}{2 T^2}
	        -\ii v \Omega }
	    }{\left(q^2-(v-\ii \epsilon )^2\right)^m}
	    \label{firstconvolutionfunction},
\end{align}
where $T$ is a positive constant, $\Omega$, $\epsilon$ and $v$ are non-negative constants, $t_{\aa}$ and $t_{\bb}$ are reaal constants, and  $m\in \{1, 2\}$.

Introducing some basic notation that we will use in this appendix,  $\mathcal{F}$ denotes the Fourier transform
\begin{align}
    \mathcal{F}\big[\,a\!\left( x \right)\big]\left(\omega\right)
    \,\coloneqq\!
    \int_{-\infty}^{\infty}
    \!\!\d x \,
    a\left(x\right)
    e^{\ii \omega x}.
\label{fouriertransform}
\end{align}
We also introduce $\ast$ to denote the convolution product, defined as
\begin{align}
    \left[a(x)\ast b(x) \right]
    \left[x\right]\,\coloneqq\! \frac{1}{2\pi}\int_{-\infty}^{\infty}\!\!\!\!\!\d \tau\,
	a(\tau)b\,(x-\tau)
	    \label{convolutionproduct}.
\end{align}

The convolution theorem allows us to write $f_m$ in \eqref{firstconvolutionfunction} as
\begin{align}
    f_m& =
    \mathcal{F}\big[\,g\!\left(v\right)\big]\!\left(\Omega\right)
    \ast
     \mathcal{F}\big[\,h_m\!\left(v\right)\big]\!\left(\Omega\right)
	    \label{convolution},
\end{align}
where $g$ and $h_m$ are functions defined as
\begin{align}
    g(v) \coloneqq
    e^{v\frac{(t_{\aa}-t_{\bb})}{T^2}
	        -\frac{v^2}{2 T^2}
	        }\,,\qquad
    h_m(v)  \coloneqq
    \frac{1
	  }{\left(q^2-(v-\ii \epsilon )^2\right)^m}\,.
\end{align}   

The Fourier transform of $g(v)$ and $h_m(v)$ are
\begin{align}
    \mathcal{F}\big[\,g\!\left( v \right)\big]\left(\Omega\right)=
    &\,\sqrt{2 \pi } T e^{\frac{\left(\ii T^2 \Omega -t_{\aa}+t_{\bb}\right)^2}{2 T^2}}
\end{align}
\begin{align}
    \mathcal{F}\big[\,h_1\!\left( v \right)\big]\left(\Omega\right)=&\,
    -\frac{
        \ii \pi  
        e^{\Omega  (\epsilon -\ii q)}
    }{2 q}
    \Big[
        \text{sgn}(\Omega ) 
        \left(-e^{2 \ii q \Omega } \text{sgn}(\left| \epsilon -\Im(q)\right| )+\text{sgn}(\left| \epsilon +\Im(q)\right| )+e^{2 \ii q \Omega }-1\right)\notag\\
    &
        -2 e^{2 \ii q \Omega }
        \text{sgn}(\epsilon -\Im(q))
        \theta (-\Omega  \text{sgn}(\epsilon -\Im(q)))
        +2 \text{sgn}(\Im(q)+\epsilon )
        \theta (-\Omega  \text{sgn}(\epsilon +\Im(q)))
    \Big]\\[4mm]
    \mathcal{F}\big[\,h_2\!\left( v \right)\big]\left(\Omega\right)=\,
    &
    \frac{
        \pi  
        e^{\Omega  (\epsilon -\ii q)}
        }{4 q^3}
        \bigg(
            \text{sgn}(\Omega )
            \Big[
                e^{2 \ii q \Omega } 
                (q \Omega +i)
                \text{sgn}(\left| \epsilon -\Im(q)\right| )
                +(q \Omega -i)
                \text{sgn}(\left| \epsilon +\Im(q)\right| )\notag\\
            &+\left(
                \ii e^{2 \ii q \Omega }
                +q \Omega 
                +q \Omega  e^{2 \ii q \Omega }
                -\ii\right)
            \Big]
            +2 \Big[
                e^{2 \ii q \Omega } 
                (q \Omega +i)
                \text{sgn}(\epsilon -\Im(q)) 
                \theta (-\Omega  \text{sgn}(\epsilon -\Im(q)))\notag\\
            &+(q \Omega -i) \text{sgn}(\Im(q)+\epsilon ) \theta (-\Omega  \text{sgn}(\epsilon +\Im(q)))
            \Big]
        \bigg).
\end{align}
Thus, using \eqref{convolution} we find that $f_m$ takes the closed forms that we use to obtain equations \eqref{notyetlabeled1} \eqref{haventlabeledit1}, i.e.,
\begin{align}
    f_1=&\,
    \left[
        \left(\text{erfi}\left(\frac{q-\ii \left(T^2 \Omega -\ii  t_{\bb} +\epsilon \right)+ t_{\aa} }{\sqrt{2} T}\right)+\ii\right)
        e^{-\frac{2 q  t_{\aa} }{T^2}+\frac{2 q  t_{\bb} }{T^2}+\frac{2 \ii q \epsilon }{T^2}+2 \ii q \Omega }+\text{erfi}\left(\frac{q+\ii \left(T^2 \Omega +\epsilon \right)- t_{\aa} + t_{\bb} }{\sqrt{2} T}\right)-\ii\right]\notag\\
    &\times         
    \left[\frac{ \pi }{2 q}
    e^{-\frac{q^2}{2 T^2}+\frac{q  t_{\aa} }{T^2}-\frac{q  t_{\bb} }{T^2}-\frac{\ii q \epsilon }{T^2}-\ii q \Omega +\frac{\ii  t_{\aa}  \epsilon }{T^2}-\frac{\ii  t_{\bb}  \epsilon }{T^2}+\frac{\epsilon ^2}{2 T^2}+\Omega  \epsilon } \right]\\[4mm]
    f_2=&\,
    \frac{ 1}{4 q^3 T^2}e^{-\frac{(q+\ii \epsilon ) \left(q+\ii \left(2 T^2 \Omega +\epsilon \right)-2  t_{\aa} +2  t_{\bb} \right)}{2 T^2}}
    \Bigg[
        \pi  \Bigg(
            -\ii q^2+q T^2 \Omega +\ii q  t_{\aa} -\ii q  t_{\bb} +q \epsilon -\ii T^2
        \notag\\&        
            e^{\frac{2 q \left(\ii \left(T^2 \Omega +\epsilon \right)- t_{\aa} + t_{\bb} \right)}{T^2}} \left(T^2+q \left(q-\ii \left(T^2 \Omega -\ii  t_{\bb} +\epsilon \right)+ t_{\aa} \right)\right) \left(\text{erfi}\left(\frac{q-\ii \left(T^2 \Omega -\ii  t_{\bb} +\epsilon \right)+ t_{\aa} }{\sqrt{2} T}\right)+\ii\right)
        \Bigg)
    \\&+
        \pi  \left(T^2+q \left(q+\ii \left(T^2 \Omega +\epsilon \right)- t_{\aa} + t_{\bb} \right)\right) \text{erfi}\left(\frac{q+\ii \left(T^2 \Omega +\epsilon \right)- t_{\aa} + t_{\bb} }{\sqrt{2} T}\right)
        -2 \sqrt{2 \pi } q T e^{\frac{\left(q+\ii \left(T^2 \Omega +\epsilon \right)- t_{\aa} + t_{\bb} \right)^2}{2 T^2}}
    \Bigg]\notag.
\end{align}

\section{Quadratic Non-local Term $\mathcal{M}$ 
\label{ap:QuadM}}
In this appendix we give the full-length closed expression of the integral over the variable $v$ in Eq. \eqref{quadMpos}, i.e., 
\begin{align}     
    \mathcal{M}_{_{t_\aa =t_\bb}}^{\op\phi^2}&=
		-\frac{\lambda ^2 
            e^{
		    -\frac{\alpha ^2}{2}
		    +2 \ii \alpha  \gamma_{\aa}
		    -\frac{ \beta ^2}{2 \delta ^2}
            }
		}{64 \pi ^4 \delta   T^3  \beta  }
		\int_0^\infty\!\!\d \xi \,
		\text{G}
		\left(\xi\right)
    \label{FinalMQuadExpr},
\end{align}
The full expression of the integrand $G(\xi)$ is 
\begin{align}    
    \text{G}
    \left(\xi\right)&\coloneqq
    \frac{1
		}{
		\xi ^2\left(\eta ^2+\xi ^2\right)}
		\sinh \left(\frac{\xi  \beta }{\delta ^2}\right)		
		e^{
		    -\frac{\xi ^2}{2 \delta ^2}
        }
        \Bigg[
            -4 \xi  \left(\sqrt{2 \pi } \eta ^2-2 \ii \eta +\sqrt{2 \pi } \xi ^2\right)
                \notag\\&\quad
            +e^{-\frac{\xi ^2}{2}} 
            \left(\eta ^2+\xi ^2\right) 
            \Bigg(
                2 \ii e^{\frac{1}{2} \eta  (\eta +2 \ii \xi )} 
                \Big(\eta  \xi  +\ii  \left(\xi ^2+1\right)\Big)   
                \bigg[
                    \text{Chi}\left(\frac{1}{2} (\eta  +\ii  \xi )^2\right) 
                \notag\\&\quad\quad\quad
                    +\ii  \pi  \,\text{erf}\left(\frac{\eta  +\ii  \xi }{\sqrt{2}}\right)
                    +2 \log (\xi -\ii \eta )
                    -\log \Big((\eta  +\ii  \xi )^2\Big)
                    -\text{Shi}\left(\frac{1}{2} (\eta  +\ii  \xi )^2\right)
                \bigg]
                \notag\\&\quad\quad
                +e^{\frac{1}{2} \eta  (\eta -2 \ii \xi )} 
                \bigg[
                    2 \left(\ii \eta  \xi +\xi ^2+1\right) \text{Chi}\left(\frac{1}{2} (\eta -\ii \xi )^2\right)
                    +2 \pi  \left(-\eta  \xi  +\ii  \left(\xi ^2+1\right)\right)
                    \,\text{erf}\left(\frac{\eta -\ii \xi }{\sqrt{2}}\right)
                    \notag\\&\quad\quad\quad
                    +4 \left(\ii \eta  \xi +\xi ^2+1\right) 
                    \log (-\xi -\ii \eta )
                    -2 \Big[\log (\xi  +\ii  \eta )+\log \Big((\eta -\ii \xi )^2\Big)\Big]
                    +\log \Big(-(\eta -\ii \xi )^2\Big)
                    \notag\\&\quad\quad\quad
                    -2 \ii \xi  (\eta -\ii \xi ) 
                    \log \Big((\eta -\ii \xi )^2\Big)
                    +2 \left(-\ii \eta  \xi -\xi ^2-1\right)
                    \text{Shi}\left(\frac{1}{2} (\eta -\ii \xi )^2\right)
                \bigg]
            \Bigg)
        \Bigg] 
    \label{finalIntegrandQuad},
\end{align}	
where Chi and Shi are the cosine and sine hyperbolic integral functions defined as
\begin{align}
    \text{Shi} (z)&\coloneqq\int _0^z\frac{ t \sinh }{t}\d t\label{Shi}\\
    \text{Chi} (z)&\coloneqq\widetilde\gamma +\int_0^z \frac{t \cosh -1}{t} \, \d t+z \log\label{Chi},
\end{align}
and $\widetilde\gamma$ here is the Euler-Mascheroni constant, and everything is expressed in terms of  dinemsnionless variables as detailed in table \ref{tab:parametersTable}.

\twocolumngrid
\bibliography{QuadraticPaperBib}
\end{document}